\documentclass[aoas,preprint]{imsart}

\RequirePackage[OT1]{fontenc}
\RequirePackage{amsthm,amsmath}
\RequirePackage[numbers]{natbib}
\RequirePackage[colorlinks,citecolor=blue,urlcolor=blue]{hyperref}

\usepackage{algorithm}
\usepackage{algpseudocode}
\usepackage{bm}
\usepackage{todonotes}

\arxiv{arXiv:0000.0000}

\startlocaldefs
\numberwithin{equation}{section}
\theoremstyle{plain}

\endlocaldefs

\begin{document}

\newcommand{\cloe}{Cloe}

\begin{frontmatter}
\title{A phylogenetic latent feature model for clonal deconvolution}
\runtitle{Phylogenetic clonal deconvolution}

\begin{aug}
\author{\fnms{Francesco} \snm{Marass}\ead[label=e1]{francesco.marass@cruk.cam.ac.uk}},
\author{\fnms{Florent} \snm{Mouliere}\ead[label=e2]{florent.mouliere@cruk.cam.ac.uk}},
\author{\fnms{Ke} \snm{Yuan}\ead[label=e3]{ke.yuan@cruk.cam.ac.uk}},
\author{\fnms{Nitzan}~\snm{Rosenfeld}\ead[label=e4]{nitzan.rosenfeld@cruk.cam.ac.uk}},
\and
\author{\fnms{Florian} \snm{Markowetz}
\ead[label=e5]{florian.markowetz@cruk.cam.ac.uk}
}

\runauthor{F. Marass et al.}

\affiliation{University of Cambridge}

\address{Cancer Research UK Cambridge Institute\\
University of Cambridge\\
Li Ka Shing Centre\\
Robinson Way\\
Cambridge, CB2 0RE\\
United Kingdom\\
\printead{e1}\\
\phantom{E-mail:\ }\printead*{e5}}
\end{aug}

\begin{abstract}
Tumours develop in an evolutionary process, in which the accumulation of mutations produces subpopulations of cells with distinct mutational profiles, called clones. This process leads to the genetic heterogeneity widely observed in tumour sequencing data, but identifying the genotypes and frequencies of the different clones is still a major challenge.
Here, we present \cloe, a phylogenetic latent feature model to deconvolute tumour sequencing data into a set of related genotypes. Our approach extends latent feature models by placing the features as nodes in a latent tree. The resulting model can capture both the acquisition and the loss of mutations, as well as episodes of convergent evolution. We establish the validity of \cloe{} on synthetic data and assess its performance on controlled biological data, comparing our reconstructions to those of several published state-of-the-art methods. We show that our method provides highly accurate reconstructions and identifies the number of clones, their genotypes and frequencies even at a modest sequencing depth. As a proof of concept we apply our model to clinical data from three cases with chronic lymphocytic leukaemia, and one case with acute myeloid leukaemia.
\end{abstract}


\end{frontmatter}


\section{Introduction}

Cancers evolve through waves of mutation and clonal expansion~\cite{Nowell1976}. Darwinian selection operates on the increased variation within the tumour, favouring clones with increased fitness, according to microenvironmental and therapeutic pressures~\cite{Fearon1990, Stratton2009, Aparicio2013, Beerenwinkel2015}. As a consequence of this evolutionary process, tumours are generally genetically heterogeneous~\cite{Gerlinger2012, Nik-Zainal2012a} and consist of related populations of cancer cells (\emph{clones}) with distinct genotypes, which encode the evolutionary history of each cell population~\cite{Nik-Zainal2012a}.
This genetic heterogeneity is important clinically because it can confound the molecular profiling of biopsies, and increased variation may equip tumours with more avenues to escape treatment, leading to worse prognosis~\cite{Schwarz2015}. 

\paragraph{The clonal deconvolution problem}
Identifying clones and their proportions is a difficult task~\cite{Beerenwinkel2015}, aggravated by the fact that cancer genomics data generally come from bulk sequencing experiments, which profile a mixture of cells from different clones. 
Clones are related to each other and can be thought of as nodes in a phylogenetic tree that describes tumour development. The root of the tree corresponds to a normal, non-mutated cell; every other node is a cancer clone with a distinct complement of mutations (its \emph{genotype}). Each clone inherits the mutations of its parent and adds more to them. This encodes a subset relationship between parent and child nodes.

\begin{figure}
  \centering
  \includegraphics[width=\textwidth]{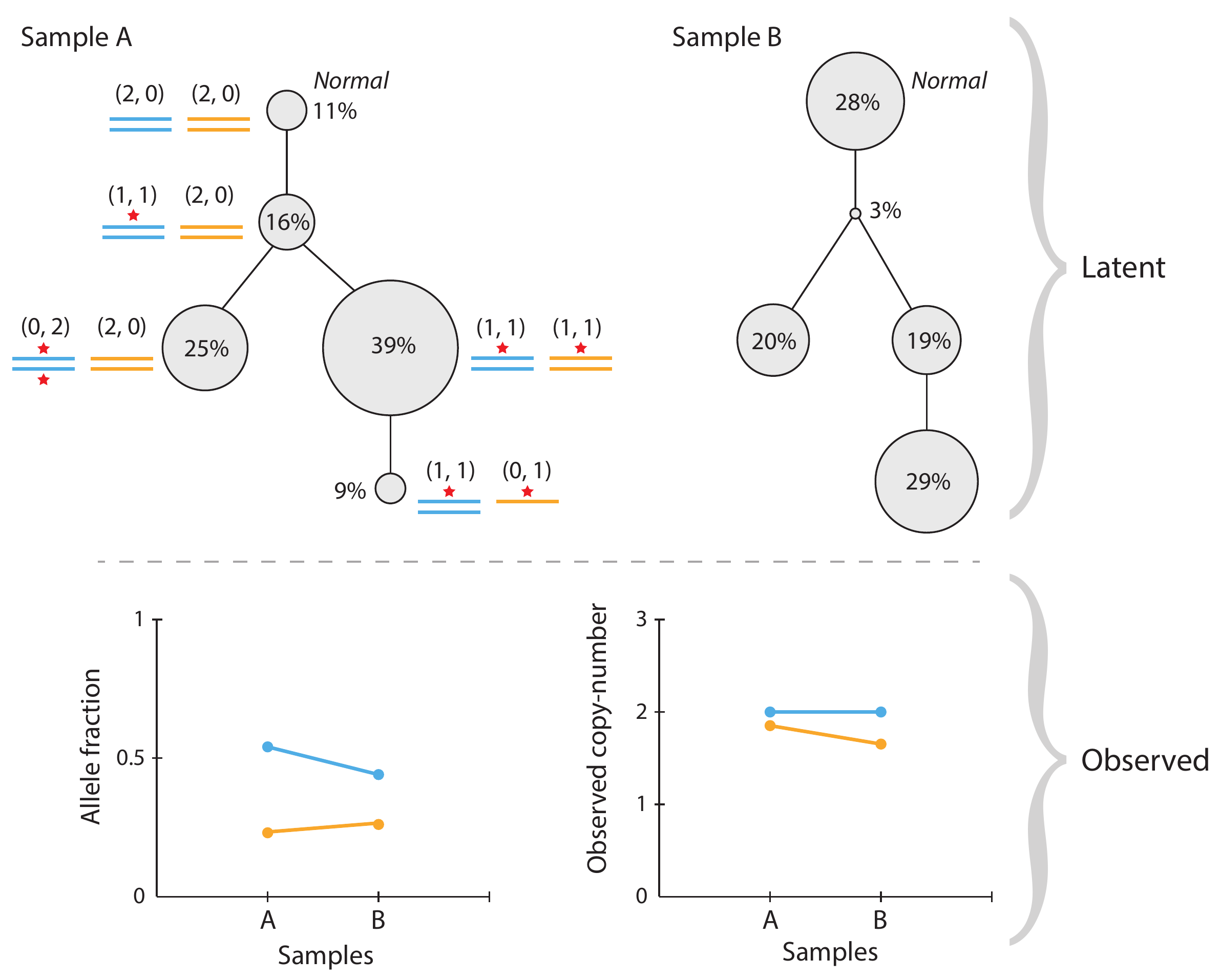}
  \caption{Overview of the general clonal deconvolution problem. Sample A shows how from a normal root node, four clones evolved according to the displayed phylogeny. In this example, genotypes consist of two loci (blue and orange), which may be mutated (red star), gained or lost. The normal genotype consists of two non-mutated alleles for each locus. Clonal fractions are represented by the diameter of the node and reported as a percentage. A second sample B from the same patient would also consist of the same tree and genotypes; clonal fractions, however, may change (as in this example). These latent parameters give rise to the observed mutation and copy-number data, shown at the bottom. The allele fraction of a mutation is the proportion of that allele in the sample. The observed copy-number is the total copy number of each clone weighted by the clonal fractions. The increase of the orange mutation's allele fraction and the decrease of its observed copy-number are due to the growth of the clone with a single mutated copy.}
  \label{fig:overview}
\end{figure}

However, none of this is directly observable. Instead, the data only consist of a set of mutations and their proportions (called \emph{allele fractions}) in a collection of tumour samples (Figure~\ref{fig:overview}).
The \emph{clonal deconvolution problem} thus asks to identify the clonal genotypes, phylogeny, and clonal fractions that best explain the observed data~\cite{ElKebir2015}.

\paragraph{Additional challenges}

The clonal deconvolution problem is further complicated by factors such as the selection of alleles during tumour evolution and the specifics of the data obtained from sequencing experiments. 
In particular, convergent evolution and mutational loss contradict the common assumption that mutations arise only once in the phylogeny (the infinite sites assumption) and never disappear. Tumours are subjected to internal selective pressures in their microenvironment and external pressures from therapeutic interventions. In such cases, multiple tumour clones may acquire the same mutation in convergent evolution, especially if it is a hotspot mutation or it confers resistance to the treatment. 
At the same time, mutations can be removed by several mechanisms, including loss of heterozygosity, the deletion of the chromosome fragment carrying the mutation.
Another challenge is that for cost-effective sequencing options like targeted amplicon sequencing, which we will use in the case studies, the depth of sequencing is not informative of the chromosomal copy-number of the tumour. This contradicts assumptions often made by previous methods.

\paragraph{Previous approaches}

Various methods have been proposed in the literature~\cite{Beerenwinkel2015} to improve on manual analyses~\cite{Gerlinger2012,Nik-Zainal2012a}.
To put our approach in context, it is useful to distinguish \emph{direct} reconstructions that directly infer clonal genotypes (like CloneHD~\cite{Fischer2014}, Clomial~\cite{Zare2014} and BayClone~\cite{Sengupta2015}) from \emph{indirect} reconstructions that obtain clusters of mutations rather than full genotypes  and require additional phylogenetic analysis to obtain clonal genotypes (like PyClone~\cite{Roth2014}, SciClone~\cite{Miller2014}, PhyloWGS~\cite{Deshwar2015}, and BitPhylogeny~\cite{Yuan2015}).

Direct reconstructions generally aim to infer two quantities, a matrix of mutation assignments and a matrix of clonal fractions, which come together in an admixture to form the sampling model. The mutation assignments matrix associates each mutation with zero or more classes, which can be intuitively interpreted as clonal genotypes. For models that lack a phylogeny, inference may yield biologically implausible genotypes, as shown later in the benchmarking studies (section~\ref{sec:benchmark}).

On the other hand, indirect methods cluster mutations based on their allele fractions across multiple samples. Joint phylogenetic modelling allows these clusters to become nodes of a tree, displaying at which node each mutation first appeared. Hence, the assignment of mutation clusters to nodes of a tree is generally inflexible to episodes of convergent evolution or mutational loss. 

\paragraph{Latent feature models}

Here we introduce \cloe, a phylogenetic latent feature model for clonal deconvolution that belongs to the category of direct reconstruction methods. 
Latent feature models discover independent features with which to describe a set of observed objects. The set of features possessed by an object determines the parameters of its distribution~\cite{Griffiths2005}. In our context, observed objects are mutations, and latent features are clonal genotypes. 

Latent feature models have been previously applied to clonal deconvolutions, but maintained the assumption that features are independent~\cite{Zare2014,Sengupta2015}.
In parallel, extensions to these models have been developed to relate features hierarchically, but placed features as the leaves of the tree~\cite{Heaukulani2014}. Moreover, these tree structure only correlated the feature assignments, making such a model unsuitable for clonal deconvolutions. 

The model we propose lifts the independence assumption and relates features with a latent hierarchy. In our framework features live at every node of the tree, thus encoding a noisy subset relationship in the mutation assignments. Our model differs from the phylogenetic Indian Buffet Process as the latter relates observed objects with a latent phylogeny, rather than the features~\cite{Miller2012}.
Our approach is more general than previously published methods, because it relies on fewer assumptions on clones and the evolutionary model: we can readily model multiple independent primary tumours, account for loss of mutations and penalise, though still allow, convergent evolution. 

We validate \cloe{} on simulated data, on a controlled biological dataset, and apply it to two published clinical datasets: longitudinal samples from three chronic lymphocytic leukaemia patients~\cite{Schuh2012}, and from an acute myeloid leukaemia case~\cite{Griffith2015}. \cloe{} is available as an R package at \url{https://bitbucket.org/fm361/cloe}.

\section{The \cloe{} model}

Our model follows the overview of Figure~\ref{fig:overview}. A latent phylogenetic tree influences the clonal genotypes; these, together with clonal fractions and additional nuisance parameters, describe the distribution of the data.

We observe data for \( J \) mutations in \( T \) samples, and the data are collected in two \( J \times T \) matrices: \textbf{X} for mutant read counts, and \textbf{D} for read depths, the number of times a particular locus of the genome is covered by sequencing reads.

The phylogenetic tree is defined by a vector \( \mathcal{T} \) with \( K > 1 \) elements, one for each clone. We consider the normal contamination as a fixed clone. Our analysis is restricted to mutations in copy-number neutral regions: each clone, including the normal, contributes exactly two copies of each allele, of which at most one can be mutated. Clonal genotypes are defined in a binary \( J \times K \) matrix \textbf{Z}, where each column \( z_{\cdot k} \) represents the genotype of clone \( k \). The proportions of each clone in each sample are summarised in a \( K \times T \) matrix \textbf{F} formed by \( T \) stochastic vectors.

Our goal is to infer the phylogeny~\( \mathcal{T} \), clonal genotypes~\textbf{Z}, and clonal fractions~\textbf{F} from the posterior distribution \( P(\mathcal{T}, \textbf{Z}, \textbf{F} \, \vert \, \textbf{X}, \textbf{D}) \). To do this, in the following sections we develop a probabilistic model that links observed and unobserved variables, and an inference algorithm to explore the posterior distribution.

\begin{figure}
  \centering
  \includegraphics[width=0.7\textwidth]{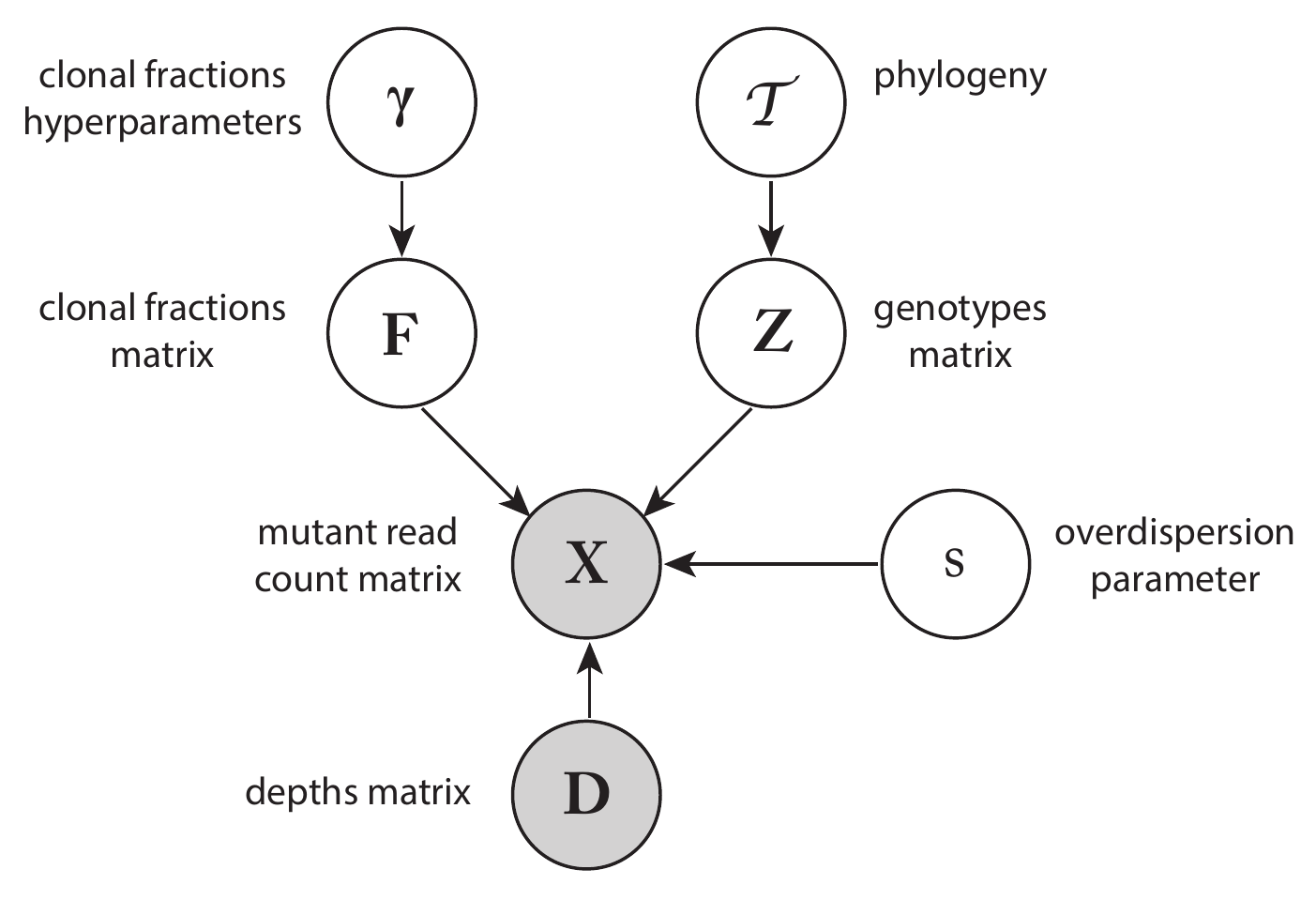}
  \caption{An outline of the graphical model corresponding to \cloe, omitting for simplicity overlapping plates and convergent evolution relations. Legend: \( \gamma \), clonal fractions hyperparameters; \textbf{F}, clonal fractions matrix; \( \mathcal{T} \), phylogeny; \textbf{Z}, genotypes matrix; \textbf{X}, mutant read counts; \textbf{D}, depths; \( s \), Beta-binomial overdispersion parameter.}
  \label{fig:graphical_model_outline}
\end{figure}

\subsection{Model definition}
\begin{figure}
  \centering
  \includegraphics[width=0.6\textwidth]{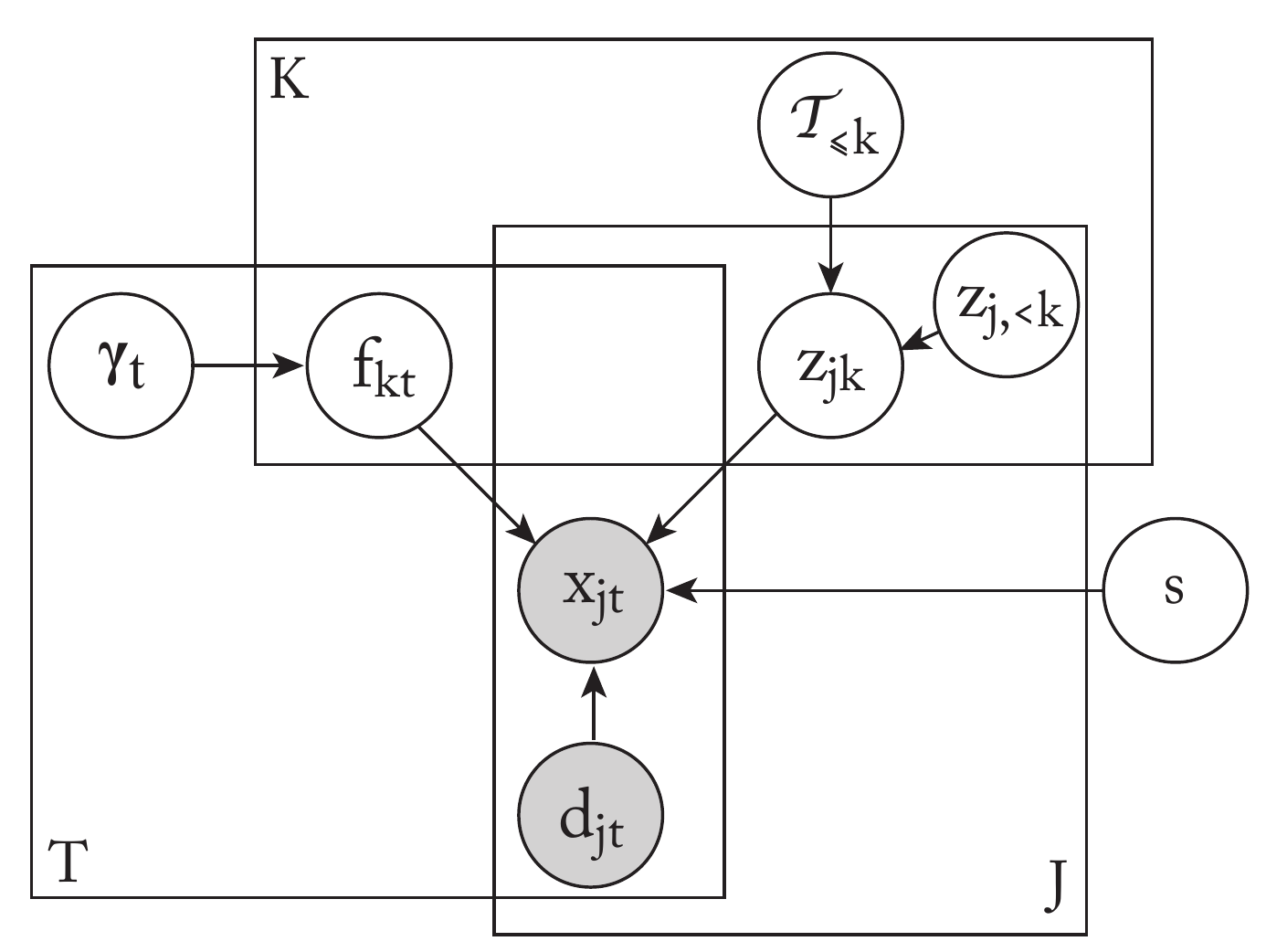}
  \caption{The full graphical model of \cloe.}
  \label{fig:graphical_model_full}
\end{figure}

For guidance, a simplified version of our model is outlined in Figure~\ref{fig:graphical_model_outline}, while at the end of this section Figure~\ref{fig:graphical_model_full} presents the complete model.

\paragraph{Phylogeny}
For \( K > 1 \) populations, we model the phylogenetic tree as a vector \( \mathcal{T} \) of length \( K \), where \( \mathcal{T}_k = l \) means that the parent of \( k \) is \( l \). The normal clone is fixed as the first entry, the root of the tree. To ensure that the graph encoded by \( \mathcal{T} \) is a tree, we let each entry only take values on the previous entries. This definition is flexible as the tree can assume any shape, even allowing phylogenies with multiple primary tumours. \( \mathcal{T} \) is defined by
\begin{align}
  \begin{aligned}
\mathcal{T}_1 & = 0, \\ 
\mathcal{T}_2 & = 1, \\
\mathcal{T}_k & \sim \mathcal{U}\left(\delta, k-1 \right) \quad\text{for } k \in \{ 3 \dots K \},
  \end{aligned}
  \label{eq:tree}
\end{align}
where \( \mathcal{U}(\delta, a) \) is a one-deflated discrete uniform distribution with values in~\( \{1, \ldots,  a  \}\). The probability of drawing a 1 is \( \delta \), and the probability of drawing an integer between \( 2 \) and \( a \) is uniform:
\begin{align}
\mathcal{U}(x; \delta, a) =
  \begin{cases}
    \delta & \text{ if } x = 1 \\
    \frac{1 - \delta}{a - 1} & \text{ if } x \in \left\{ 2, 3, \dots, a \right\}
  \end{cases}
\end{align}
We penalise multiple independent primary tumours (multiple children of the normal clone) by setting \( \delta = (2 k)^{-1} \).

\paragraph{Genotypes}
Genotypes are defined in a binary \( J \times K \) matrix \(\text{\textbf{Z}} = (z_{jk}) \), where \(1\) denotes a mutation and \(0\) the un-mutated (\emph{wildtype}) state, for each mutation \( j \) in each clone \( k \). 
We fix the genotype of the normal clone to a zero vector of length \( J \), implying that all mutations are somatic. More generally, the normal genotype could be modified to accommodate for known germline variants. The genotype of a clone \( k \) for a mutation \( j \) is then defined as 
\begin{equation}
z_{jk} \, \vert \, z_{j\mathcal{T}_k}, \mu, \rho \sim \text{Bernoulli}(p_{jk})
\label{eq:genotypes}
\end{equation}
where \( \mu \) is the probability of mutating if the parent does not have a mutation, and
\( \rho \) is the probability of reverting to wildtype if the parent is mutated:
\begin{equation}
p_{jk} =
\begin{cases}
\,\mu  & \text{ if } z_{j\mathcal{T}_k} = 0 \\
\,1 - \rho & \text{ if } z_{j\mathcal{T}_k} = 1.
\end{cases}
\label{eq:genotypes_probability}
\end{equation}

\paragraph{Clonal fractions}
Because the samples may be collected latitudinally, longitudinally, and at irregular intervals, we assume that clonal fractions are independent between samples. We represent clonal fractions with a \( K \times T \) matrix \( \mathbf{F} = (\mathbf{f}_{\cdot t})_{t = 1, \ldots, T} \) composed of a  collation of stochastic column vectors \( \mathbf{f}_{\cdot t} \) describing the proportions of each clone in a sample \(t\). Clonal fractions for a sample \( t \) are modelled with a symmetric Dirichlet distribution with hyperparameter \( \gamma_t \):
\begin{equation}
\mathbf{f}_{\cdot t} \, \vert \, \gamma_t \sim \text{Dirichlet}(\gamma_t)
\end{equation}

\paragraph{Likelihood}
Genotypes and clonal fractions come together in an admixture, their dot product representing the expected allele fractions for each mutation in each sample. 
We model mutant reads as successful trials from a beta-binomial distribution with overdispersion parameter \( s \). The probability of success is a function of the expected allele fraction \( p_{jt} = \frac{1}{2} \left( \mathbf{z}_{j \cdot} \cdot \mathbf{f}_{\cdot t} \right) \). To capture sequencing noise at extreme values of \( p_{jt} \) we replace it with a function \( e(p_{jt}) \) that depends on the sequencing error rate \( \varepsilon \) (e.g. 0.1\%) such that
\begin{equation}
e(p_{jt}) \; =
\begin{cases}
\; \varepsilon & \text{ if } p_{jt} = 0 \\
\; 1 - \varepsilon & \text{ if } p_{jt} = 1 \\
\; p_{jt} & \text{ otherwise.}
\end{cases}
\label{eq:seq_noise}
\end{equation}

The likelihood is then specified by
\begin{equation}
\label{eq:cnn_likelihood}
x_{jt} \, \vert \, d_{jt}, \mathbf{z}_{j \cdot}, \mathbf{f}_{\cdot t}, s \;\sim\; \text{Beta-binomial}(d_{jt}, e(p_{jt}), s).
\end{equation}

\paragraph{Nuisance parameters}
We let the beta-binomial overdispersion parameter \( s \) and the Dirichlet hyperparameters \( \bm{\gamma} \) have Gamma priors, whereas the mutation and reversion probabilities \( \mu \) and \( \rho \) are fixed.

\subsection{Penalising convergent evolution}
One of the risks of assuming the independence of features in this biological application is that the inferred genotypes may largely display convergent evolution. We can penalise such occurrences by altering the definition of genotypes (cfr. equations~\ref{eq:genotypes} and \ref{eq:genotypes_probability}).

Under the infinite sites assumption (ISA) every mutation occurs only once, so that if multiple clones possess a mutation \( j \), then the mutation must have appeared with their most recent common ancestor. In contrast, if the most recent common ancestor is not mutated, then the mutation must have appeared multiple times (convergent evolution).
We thus say that a mutation assignment \( z_{jk} = 1 \) conflicts with ISA if the most recent common ancestor of \( \left\{ k' : z_{jk'} = 1, k' \leq k \right\} \) does not harbour mutation \( j \). 

We include ISA checks into our model by using an indicator function \( I(j, k, a) \) that returns \( 1 \) if the most recent common ancestor of all clones \( k' \leq k \) that harbour mutation \( j \) also possesses the mutation, when \( z_{jk} = a \).
That is, \( I(j, k, a) = 1 \) if ISA is satisfied by setting \( z_{jk} = a \), and 0 otherwise.  

Thus we redefine the distribution of genotypes making them conditional on all previous genotypes and weighting assignments by a user-defined parameter \( \nu \) if they comply with ISA, or by \( 1 - \nu \) if they do not:
\begin{equation}
P(z_{jk} = 1 \, \vert \, \mathcal{T}, \mathbf{Z}_{j, < k}, z_{j\mathcal{T}_k} = 0, \mu, \rho, \nu) \propto ( \mu \nu )^{I(j, k, 1)} ( \mu ( 1 - \nu ) )^{1 - I(j, k, 1)}
\label{eq:genotypes_isa_probability}
\end{equation}
In practice, only transitions that gain a mutation can clash with ISA, and the factor of \( \nu \) immediately cancels out if the parental genotype is 1 at a given \( j \). An ISA-check at \( j \) is thus only warranted if the parent is not mutated at \( j \).

The graphical model corresponding to what has been described so far is shown in Figure~\ref{fig:graphical_model_full}.

\begin{algorithm}
  \begin{algorithmic}[1]
    \For{\( i \) = 1, \dots, \#\text{iterations}}
      \For{\( m \) = 1, \dots, \#\text{chains}}
        \For{\( j \) = 1 \dots \( J \)}
        \Comment \textbf{Z}
          \State Propose new \( \mathbf{z}^{* (m)}_{j \cdot} \)
          \State Accept with probability \ref{eq:inference_z_mh_ratio}
        \EndFor
        \For{\( k \) = 3 \dots \( K \)}
          \Comment \( \mathcal{T} \)
          \State Compute \( P(\mathcal{T}^{*(m)}_k = l) \) for \( l \in \left\{ 1 \dots k - 1 \right\} \) (eq.~\ref{eq:inference_tree})
          \State Sample new \( \mathcal{T}^{*(m)}_k \) from \( P(\mathcal{T}^{*(m)}_k) \)
        \EndFor
        \State Randomly swap two siblings
        \State With probability 1\% propose a swap between a node and its parent
        \State Accept with probability~\ref{eq:parent_swap}
        \For{\( t \) = 1 \dots \( T \)}
          \Comment \textbf{F}
          \State Propose new \( \mathbf{f}^{*(m)}_{\cdot t} \) from eq.~\ref{eq:inference_f_proposal}
          \State Accept with probability \ref{eq:inference_f_mh_ratio}
        \EndFor
        \Comment Nuisance parameters
        \State Propose new \( s^{*(m)} \sim \mathcal{N}(s^{(m)}, \sigma_s) \) and accept with probability~\ref{eq:inference_s_mh_ratio}
        \For{\( t \) = 1 \dots \( T \)}
          \State Propose new \( \gamma^{*(m)}_t \sim \mathcal{N}(\gamma^{(m)}_t, \sigma_\gamma) \) and accept with probability~\ref{eq:inference_gamma_mh_ratio}
        \EndFor
      \EndFor
      \If{\( i \) is a multiple of 100}
        \Comment Chain swap
        \State Propose a chain \( j \in \{ 1 \dots m - 1 \} \)
        \State Accept the state swap between chains \( j \) and \( j + 1 \) with probability~\ref{eq:inference_swap_mh_ratio}
      \EndIf
    \EndFor
  \end{algorithmic}
  \caption{MCMCMC sampling algorithm for \cloe}
  \label{alg:inference}
\end{algorithm}


\subsection{Inference}

We are interested in the posterior distribution of the latent variables given the observed variables
\(P(\mathcal{T}, \mathbf{Z}, \mathbf{F} \, \vert \, \mathbf{X}, \mathbf{D})\),
which we explore by Metropolis-coupled Markov chain Monte Carlo (MCMCMC)~\cite{Geyer1991}, integrating out the nuisance parameters \( s\) and \( \bm{\gamma} \).
We use Gibbs sampling to update the tree vector \( \mathcal{T} \) and Metropolis-Hastings moves to update the other quantities.

Because the posterior landscape appears composed of high peaks separated by deep valleys (Supplementary Figure 1), we run five chains in parallel, with tempered posteriors. 
The sampling strategy described hereafter is applied to each chain, and summarised in Algorithm~\ref{alg:inference}.

\paragraph{Phylogeny}
For each \( \mathcal{T}_{k > 2} \), we compute the conditional posterior of the parent assignment \( \mathcal{T}_k = l \), for each \( l < k \):
\begin{equation}
P( \mathcal{T}_k = l \, \vert \, \dots) \propto P(\mathbf{Z} \, \vert \, \mathcal{T}_{-k}, \mathcal{T}_k = l) \,
P(\mathcal{T}_k = l)
\label{eq:inference_tree}
\end{equation}
The likelihood term amounts to tallying the genotype transitions from parent \( l \) to child \( k \), and reassessing how many transitions comply or clash with ISA for all clones. The prior, according to eq.~\ref{eq:tree}, is equal to
\begin{equation}
  \begin{aligned}
P(\mathcal{T}_k = l) & = \delta^{\bm{I}(l = 1)} \left( \frac{1 - \delta}{k - 2} \right)^{1 - \bm{I}(l = 1)} \\
& = \left( \frac{1}{2k} \right)^{\bm{I}(l = 1)} \left( \frac{2k - 1}{2k \, (k - 2)} \right)^{1 - \bm{I}(l = 1)}
  \end{aligned}
  \label{eq:inference_tree_prior}
\end{equation}

To facilitate the exploration of the space of tree and genotypes configurations, we uniformly propose a pair of siblings and swap their position in the tree and in the genotypes and clonal fractions matrices. Prior to this swap, the siblings had access to one linear topology. This move allows the other linear topology to be explored, while leaving probabilities unaltered.

In addition, the swap between a node \( k \) and its parent \( l \) is proposed. A node \( k \) is chosen uniformly from \( \left\{ 3 \dots K \right\} \). A tree \( \mathcal{T}^* \) is created where \( k \) is the parent of \( l \), while any children of \( k \) remain children of \( k \); the same applies to \( l \). As with the sibling swap, this move requires rearranging the clone order in the genotypes and clonal fractions matrices. The parent swap affects genotype transitions from \( \mathcal{T}_l \), the original parent of \( l \), to \( k \), and from \( k \) to \( l \). The proposal is accepted with probability
\begin{equation}
\min\left( 1, \frac{P(\mathbf{Z}_{\{ \mathcal{T}_l, k, l \}} \, \vert \, \mathcal{T}^*, \mu, \rho, \nu) \, P(\mathcal{T}^*_{\{ \mathcal{T}_l, k, l \}})}{P(\mathbf{Z}_{\{ \mathcal{T}_l, k, l \}} \, \vert \, \mathcal{T}, \mu, \rho, \nu) \, P(\mathcal{T}_{\{ \mathcal{T}_l, k, l \}})} \right)
\label{eq:parent_swap}
\end{equation}
We perform this update with probability 0.01.

\paragraph{Genotypes}
Because mutations are independent, we update \textbf{Z} by row, proposing a new row \( \mathbf{z}^*_{j \cdot} \) by flipping each bit of \( \mathbf{z}_{j \cdot} \) with probability \( \theta \). The proposal is symmetric and the move is accepted with probability
\begin{equation}
\min\left( 1,  \frac{P(\mathbf{x}_{j \cdot} \, \vert \, \mathbf{d}_{j \cdot}, \mathbf{z}^*_{j \cdot}, \mathbf{F}, s) \, P(\mathbf{z}^*_{j \cdot} \, \vert \, \mathcal{T}, \mu, \rho, \nu)}{P(\mathbf{x}_{j \cdot} \, \vert \, \mathbf{d}_{j \cdot}, \mathbf{z}_{j \cdot}, \mathbf{F}, s) \, P(\mathbf{z}_{j \cdot} \, \vert \, \mathcal{T}, \mu, \rho, \nu)} \right)
\label{eq:inference_z_mh_ratio}
\end{equation}
where the likelihood is only computed for mutation \( j \), and the prior refers to the sequence of transitions from the root genotype at \( j \) to the leaves, with appropriate penalties for convergent evolution.

\paragraph{Clonal fractions}
Because of the independence of the samples, the matrix \textbf{F} is updated by column. A new vector \( \mathbf{f}^*_{\cdot t} \) is proposed from a Dirichlet distribution centred at the current value \( \mathbf{f}_{\cdot t} \):
\begin{equation}
Q(\mathbf{f}^*_{\cdot t} \, \vert \, \mathbf{f}_{\cdot t}) = \text{Dirichlet}(\psi \, \mathbf{f}_{\cdot t} + \epsilon)
\label{eq:inference_f_proposal}
\end{equation}
where \( \psi \) is a precision factor and \( \epsilon \) a small bias to avoid sinks at 0. The proposal is accepted with probability
\begin{equation}
\min\left( 1, \frac{P(\mathbf{x}_{\cdot t} \, \vert \, \mathbf{d}_{\cdot t}, \mathbf{Z}, \mathbf{f}^*_{\cdot t}, s) \ P(\mathbf{f}^*_{\cdot t} \, \vert \, \gamma_t) \ Q(\mathbf{f}_{\cdot t} \, \vert \, \mathbf{f}^*_{\cdot t})}{P(\mathbf{x}_{\cdot t} \, \vert \, \mathbf{d}_{\cdot t}, \mathbf{Z}, \mathbf{f}_{\cdot t}, s) \ P(\mathbf{f}_{\cdot t} \, \vert \, \gamma_t) \ Q(\mathbf{f}^*_{\cdot t} \, \vert \, \mathbf{f}_{\cdot t})} \right)
\label{eq:inference_f_mh_ratio}
\end{equation}

\paragraph{Nuisance parameters}

The remaining parameters are updated with Metropolis moves using Gaussian proposals. The Metropolis-Hastings acceptance ratios are
\begin{equation}
\min\left( 1, \frac{P(\mathbf{X} \, \vert \, \mathbf{D}, \mathbf{Z}, \mathbf{F}, s^*) \ P(s^*)}{P(\mathbf{X} \, \vert \, \mathbf{D}, \mathbf{Z}, \mathbf{F}, s) \ P(s)} \right) \quad\text{for}~s,
\label{eq:inference_s_mh_ratio}
\end{equation}
and
\begin{equation}
\min\left( 1, \frac{P(\mathbf{f}_{\cdot t} \, \vert \, \gamma^*_t) \ P(\gamma^*_t)}{P(\mathbf{f}_{\cdot t} \, \vert \, \gamma_t) \ P(\gamma_t)} \right) \quad\text{for}~\gamma_t.
\label{eq:inference_gamma_mh_ratio}
\end{equation}

\paragraph{Temperatures and chain swaps}

Regularly at user-defined intervals, a swap between two adjacent chains is proposed as a Metropolis-Hastings move~\cite{Geyer1991}. Let \( M \) denote the number of parallel chains, \( P^{(m)} \) denote the tempered posterior of chain \( m \), and \( \omega_m \) denote the state of chain \( m \). A chain \( m \) is selected among the first \( M - 1 \) chains. The swap between chains \( m \) and \( m + 1 \) is then accepted with probability
\begin{equation}
\min\left( 1, \frac{P^{(m)}(\omega_{m+1}) \ P^{(m+1)}(\omega_m)}{P^{(m)}(\omega_{m}) \ P^{(m+1)}(\omega_{m+1})} \right).
\label{eq:inference_swap_mh_ratio}
\end{equation}

The temperature \( \tau_m \) for each chain \( m \) is chosen according to the following scheme:
\begin{equation}
\tau_m = (1 + \Delta T (m - 1))^{-1}
\end{equation}
where \( \Delta T > 0 \) regulates the temperature differences between chains.

\paragraph{Parameter estimates}

MCMCMC parameter estimates are derived solely from the first, untempered chain. After discarding a certain proportion of the initial samples as burn-in, and thinning the chain by a factor of \( i \), thus considering every \( i^{th} \) sample, we obtain a maximum \emph{a posteriori} (MAP) estimate of the parameters by selecting the chain state of the sample with the highest posterior value.

\section{Validation and benchmarks}

We extensively validated and benchmarked \cloe{} by using simulated data and a controlled experimental set-up based on mixing cell lines. 

\subsection{Simulated data}

We first tested our model on 9 simulated datasets, one for each combination of number of clones (3, 4 or 5) and depth of sequencing (means: 50\( \times \), 200\( \times \), 1000\( \times \)).
The genotypes were created according to a random tree and using parameters \( \mu = 0.5 \), \( \rho = 0.05 \), \( \nu = 0.9 \); clonal fractions were \emph{iid} draws from a symmetric Dirichlet distribution with parameter \( \gamma = 2 \). All datasets contained 100 mutations and 5 samples, with depths obtained from a Poisson distribution and mutant read counts obtained from a binomial distribution. Because of the fixed size of our model, we ran \cloe{} for 3, 4, and 5 clones on each of the 9 datasets (running parameters are reported in Table~\ref{tab:params}).

We measured \cloe's performance in several ways: we evaluated its ability to identify the right number of clones, we assessed mixing by computing the Gelman-Rubin statistic from three consecutive runs of the algorithm, and finally we measured the reconstruction error. To perform the latter step, we calculated two metrics, the normalised genotypes error \( Z_{err} \) and the normalised clonal fractions error \( F_{err} \), both defined as the sum of the absolute differences between inferred (\( \mathbf{Z}^* \)) and the real (\( \mathbf{Z} \)) matrices, normalised by the real matrix dimensions (ignoring the fixed genotype in \textbf{Z}).
To control for equivalent solutions with permuted clones, we used permutations \(\sigma\) of the columns of \( \mathbf{Z}^* \) to minimise \( Z_{err} \): 
\begin{equation}
Z_{err} = \frac{1}{J (K - 1)} \, \min_{\sigma} \left(\sum_{j,k} \vert z^*_{j\sigma(k)} - z_{jk} \vert \right),
\end{equation}
 The same permutation is then used to rearrange the rows of \( \mathbf{F}^* \), which is normalised by \( J K \). If the real and inferred matrices have different sizes, we pad the smaller one with normal clones with zero clonal fractions. In this instance \( K \) refers to the real number of clones. 

\begin{figure}
  \centering
  \includegraphics[width=\textwidth]{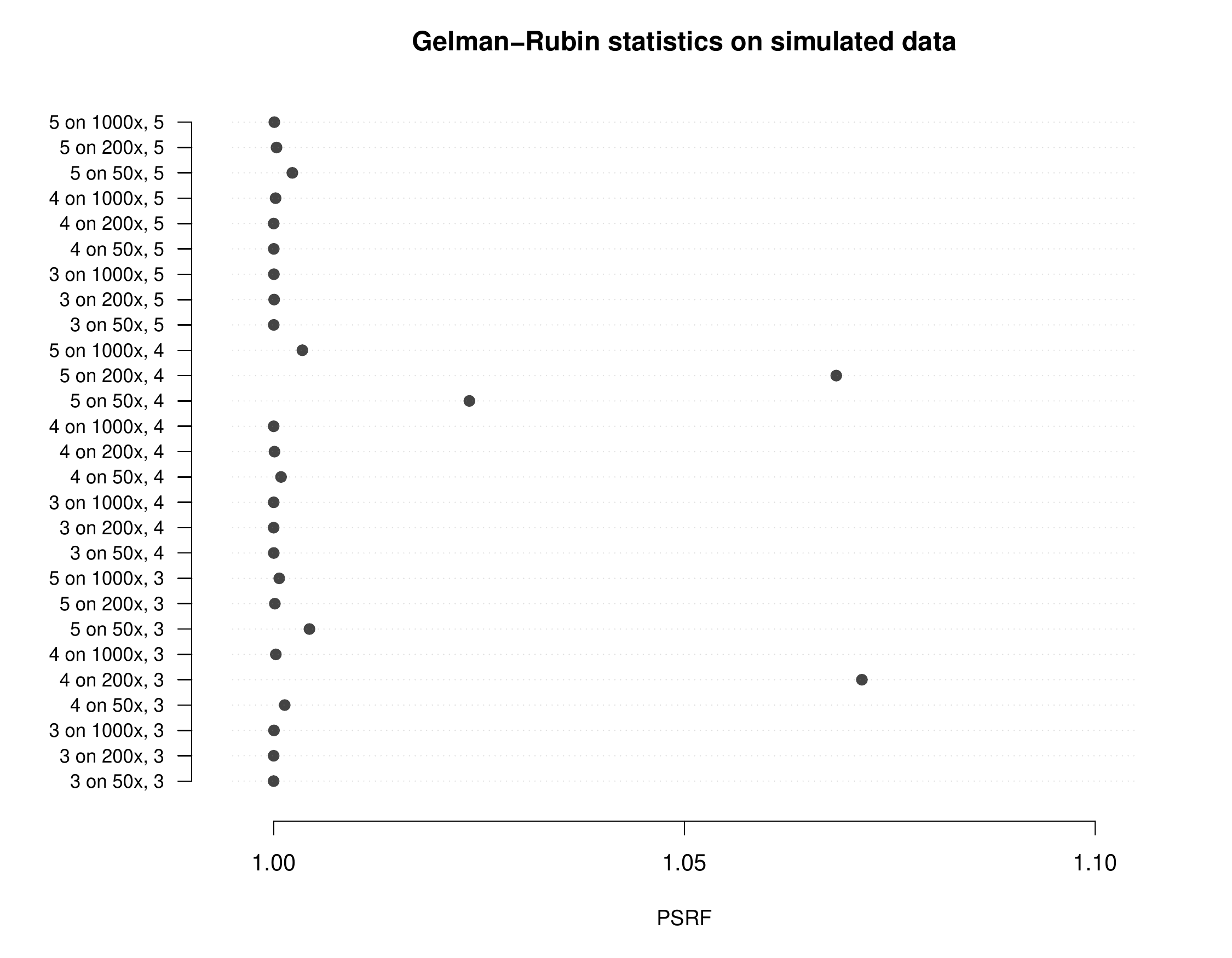}
  \caption{The Gelman-Rubin statistic computed from three runs of our algorithm on the simulated datasets. Row names consist of the number of clones that was used in the run, and the characteristics of the dataset (sequencing depth and the real number of clones).}
  \label{fig:simulation_psrf}
\end{figure}

\paragraph{MCMC mixing and effective sample size}
The Gelman--Rubin statistic was calculated from the log-posterior values of the untempered chains as a proxy for the multidimensional parameters. In every case, the potential scale reduction factor is within the accepted range, less than 1.1 (Figure~\ref{fig:simulation_psrf}). Each run was started with a different random seed.

We computed the effective sample size (ESS) for the first run on each dataset, focussing on the cases where the sought number of clones matched the real number of clones of the dataset. Table~\ref{tab:ess} reports the ESS of the nuisance parameters and the log-posterior computed from 10,000 post-thinning iterations. Modulating the standard deviation of the Gaussian proposals for the nuisance parameters could decrease their autocorrelation. However, the shape of the posterior space~(Supplementary Figure 1) may prevent efficient large moves.

\begin{table}
  \centering
  \begin{tabular}{lrrrrrrr}
  \hline
Dataset & \( \gamma_1 \) & \( \gamma_2 \) & \( \gamma_3 \) & \( \gamma_4 \) & \( \gamma_5 \) & \( s \) & LP \\
  \hline
50x, 3   & 194.63 & 272.20 & 474.59 & 117.33 & 134.96 & 1696.31 & 4366.40 \\
200x, 3  & 178.25 & 352.60 & 505.93 & 171.06 & 151.80 & 890.28 & 3088.45 \\
1000x, 3 & 143.95 & 442.33 & 446.48 & 220.48 & 153.46 & 426.54 & 1685.87 \\
50x, 4   & 648.35 & 241.09 & 586.84 & 170.57 & 342.17 & 1345.90 & 2323.20 \\
200x, 4  & 735.31 & 195.13 & 683.51 & 152.63 & 346.05 & 1025.92 & 1700.03 \\
1000x, 4 & 1083.17 & 302.75 & 654.13 & 140.00 & 237.40 & 406.45 & 894.08 \\
50x, 5   & 395.45 & 239.42 & 209.06 & 261.33 & 369.81 & 1182.60 & 1673.17 \\
200x, 5  & 803.71 & 387.01 & 394.54 & 324.07 & 644.47 & 947.42 & 2401.41 \\
1000x, 5 & 990.07 & 401.38 & 373.32 & 284.74 & 657.88 & 478.26 & 1805.82 \\
   \hline
  \end{tabular}
  \caption{Effective sample size per dataset. The effective sample size was computed on the 10,000 post-thinning iterations for the first of the three replicate runs with the correct number of clones. The dataset is denoted by the average sequencing depth and the real number of clones. LP denotes the log-posterior, as a proxy for the multidimensional parameters.}
  \label{tab:ess}
\end{table}

\begin{figure}
  \centering
  \makebox[\textwidth][c]{
    \includegraphics[scale=0.52]{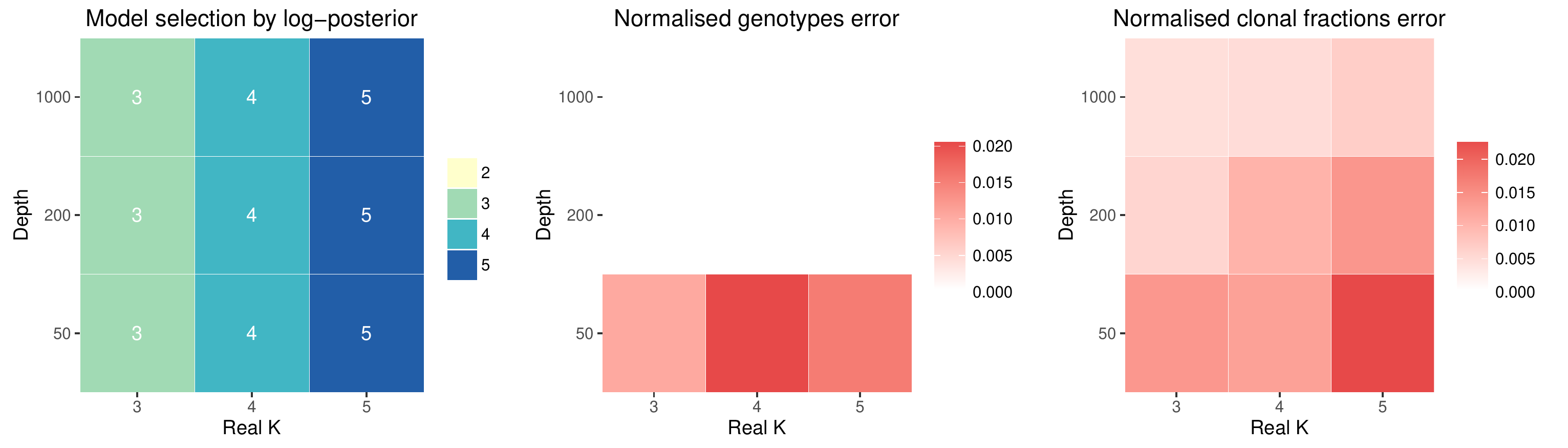}
  }
  \caption{Results on the simulated datasets. Left: inferred model size for every combinations of \( K \) and depths. Centre and right: reconstruction errors. All datasets consisted of five samples and 100 mutations.}
  \label{fig:simulation_results}
\end{figure}

\paragraph{Model selection}
As a model selection criterion we used the log-posterior probability of the MAP sample. We were able to recover the correct size \( K \) for every dataset (Figure~\ref{fig:simulation_results}, left) in two of the three runs. In one run, a higher log-posterior probability was given to the solution with 4 clones on the dataset with 50\( \times \) and 5 clones (-14464.05 compared to the 5-clone solution's -14464.58; Supplementary Figure 2). In such cases, where the log-posterior probabilities of different models are approximately equal, we prefer the solution with higher log-likelihood value.

\paragraph{Reconstruction fidelity}
To assess our reconstructions we considered only the MAP solution for each dataset. The reconstruction error was low, with \( Z_{err} \leq 0.027 \) and \( F_{err} \leq 0.033 \). The largest errors were obtained at the lowest depth (Figure~\ref{fig:simulation_results}, centre and right), suggesting that on these random datasets \cloe{} can not only discover the correct number of clones,
but also infer correct genotypes and clonal fractions with \( > 96.7\% \) accuracy (Supplementary Figures 3 and 4).

\paragraph{Conclusion}
This validation with synthetic data provided proof that the model and the inference are sound, achieving good reconstructions even with several clones and low sequencing depth (Supplementary Figure 5).

\begin{table*}
  \begin{tabular}{lr}
\hline
\textbf{Parameter} & \textbf{Value} \\
\hline
\multicolumn{2}{c}{MCMCMC} \\
\hline
iterations & 40000 \\
chains & 5 \\
\( \Delta T \) & 0.4 \\
swap interval & 50 \\
burn-in & 50\% \\
thinning factor & 4 \\
\hline
\multicolumn{2}{c}{\textbf{Z}} \\
\hline
\( \mu \) & 0.3 \\
\( \rho \) & 0.1 \\
\( \nu \) & 0.75 \\
\( \theta \) (proposal) & 0.20 \\
\( \varepsilon \) (likelihood) & 0.005, 0.002 \\
\hline
\multicolumn{2}{c}{\textbf{F}} \\
\hline
\( \psi \) (proposal) & 200 \\
\( \epsilon \) (proposal) & 4 \\
\hline
\multicolumn{2}{c}{Nuisance parameters} \\
\hline
\( \gamma \) (prior, shape) & 2 \\
\( \gamma \) (prior, rate) & 1 \\
\( \sigma_\gamma \) (proposal) & 0.2 \\
\( s \) (prior, shape) & 11 \\
\( s \) (prior, rate) & 0.10 \\
\( \sigma_s \) (proposal) & 16 \\
\hline
  \end{tabular}
  \caption{Running parameters for the simulated and validation datasets. For \( \varepsilon \), the sequencing error parameter, the first value refers to the simulations, the second to the validation dataset.}
  \label{tab:params}
\end{table*}


\subsection{Controlled experimental data}

Because synthetic data may not capture the variability seen in real biological data, we tested our method on a bespoke experiment.
In order to mimic heterogeneous tumour samples, we genotyped five cell lines and mixed them together at known proportions. Four single-cell-diluted cancer cell lines (HD124, HD212, HD249, and HD659, from Horizon Discovery, UK) were mixed with a normal cell line (HG00131, 1000 Genomes Project, Coriell Institute, USA). All five cell lines were subjected to whole-exome sequencing (Nextera Rapid Capture Exome, Illumina, USA).
Mutational and single nucleotide polymorphism (SNP) profiles were obtained using the standard samtools workflow \cite{samtools}. To identify copy-number neutral regions, we generated copy-number profiles with the R package CopywriteR (version 2.2.0). We created 14 mixtures, with a median tumour cellularity of 64\% (Table~\ref{tab:mixtures}).

\begin{figure}
  \centering
  \makebox[\textwidth][c]{
    \includegraphics[scale=0.5]{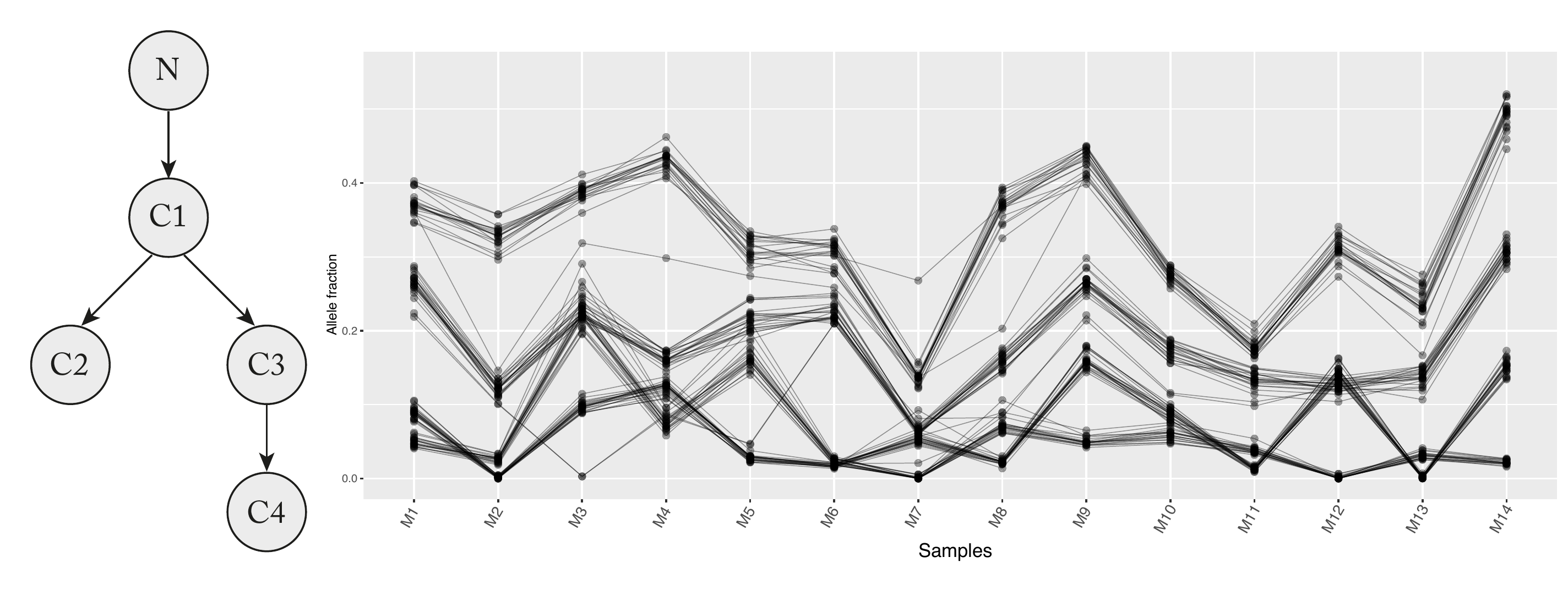}
  }
  \caption{The observed data for the validation dataset. On the left is the artificial phylogeny, where N denotes the normal clone, C1 to C4 are the cancer cell lines, playing in this context the role of clones. On the right are the observed mutational dynamics (allele fractions over samples) at an average depth of 17260\( \times \).}
  \label{fig:data_cnn}
\end{figure}

Because the cell lines were unrelated, we selected a subset of mutations (heterozygous single nucleotide variants and small indels) so as to embed the cell lines into an artificial phylogeny (Figure~\ref{fig:data_cnn}, left). We focussed on regions that were copy-number neutral across all cell lines, and measured allele fractions by targeted sequencing (Figure~\ref{fig:data_cnn}, right) \cite{Forshew2012}. We excluded from the final dataset mutations whose genotypes had been erroneously inferred from exome sequencing data. The final dataset included 82 mutations, of which one displayed convergent evolution and two were reversions to wildtype, and was sequenced to a median of 17260\( \times \) coverage. 
Targeted sequencing data were processed with an \emph{in house} pipeline and the known mutations were quantified from pileups with a base quality cutoff of 30. 

\begin{table*}
  \centering
  \makebox[\textwidth][c]{
    \begin{tabular}{rrrrrrrrrrrrrrr}
\hline
\ & M1 & M2 & M3 & M4 & M5 & M6 & M7 & M8 & M9 & M10 & M11 & M12 & M13 & M14 \\
\hline
N  & 0.26 & 0.34 & 0.22 & 0.13 & 0.37 & 0.38 & 0.72 & 0.26 & 0.13 & 0.45 & 0.65 & 0.38 & 0.53 & 0.00 \\
C1 & 0.03 & 0.41 & 0.14 & 0.29 & 0.14 & 0.14 & 0.04 & 0.38 & 0.03 & 0.02 & 0.06 & 0.09 & 0.19 & 0.08 \\
C2 & 0.18 & 0.00 & 0.19 & 0.25 & 0.06 & 0.04 & 0.11 & 0.05 & 0.32 & 0.18 & 0.03 & 0.28 & 0.00 & 0.30 \\
C3 & 0.44 & 0.19 & 0.00 & 0.17 & 0.10 & 0.41 & 0.13 & 0.18 & 0.43 & 0.23 & 0.20 & 0.25 & 0.22 & 0.57 \\
C4 & 0.10 & 0.05 & 0.44 & 0.15 & 0.33 & 0.05 & 0.00 & 0.14 & 0.10 & 0.12 & 0.07 & 0.00 & 0.06 & 0.04 \\
\hline
    \end{tabular}
  }
  \caption{Clonal fractions in the 14 mixtures of the validation experiment.}
  \label{tab:mixtures}
\end{table*}

The large number of samples and the high depth of sequencing that we obtained afforded a sensitivity analysis, in which we varied the number of samples and the depth. \cloe{} was run on these datasets with the same parameters as for the synthetic data (Table~\ref{tab:params}).

\paragraph{Model selection}

\begin{figure}
  \centering
  \includegraphics[scale=0.75]{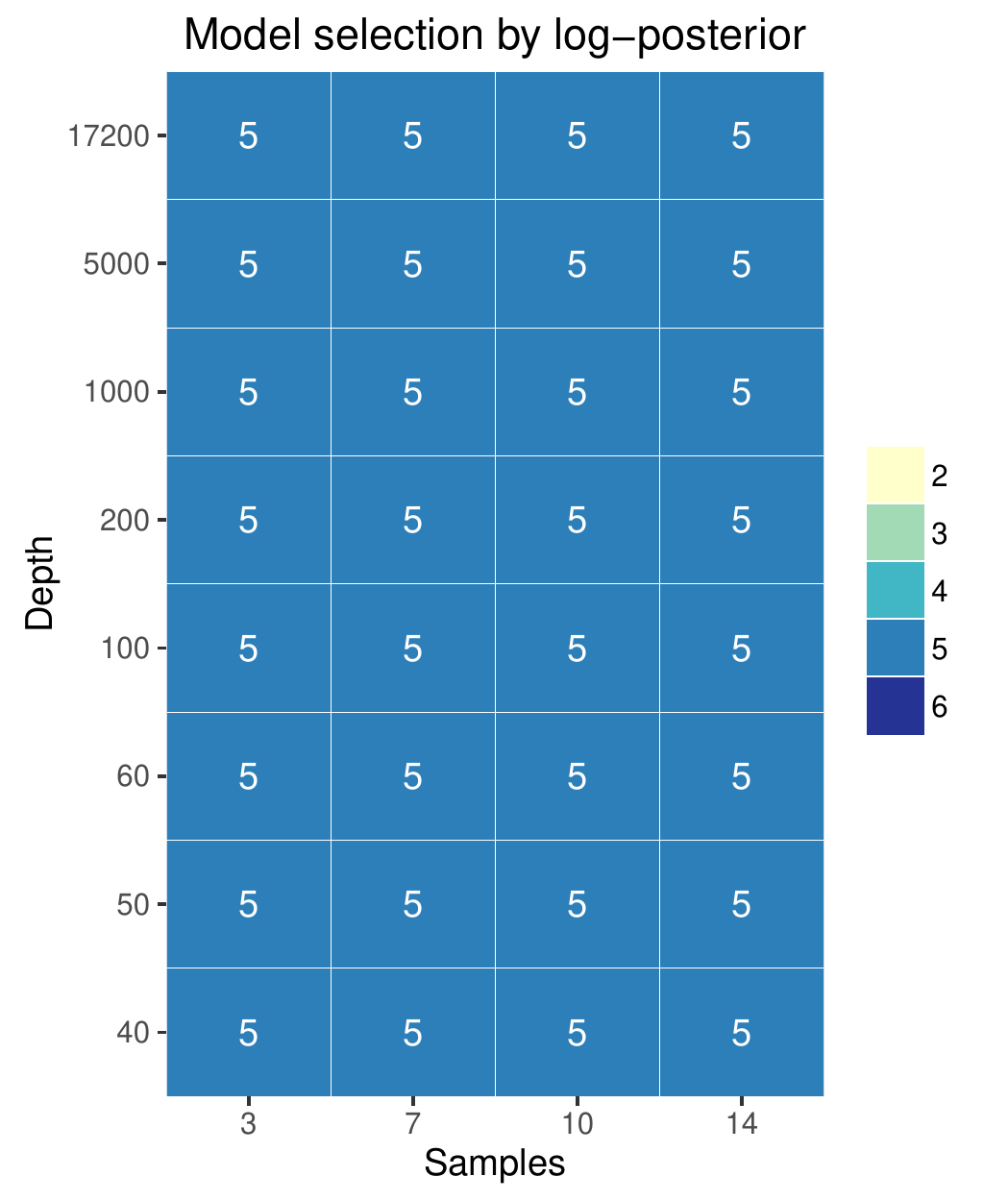}
  \caption{Inferred model size \( K \) for every combination of samples and depths in the validation dataset. The validation data consist of a mixture of five clones. Only the top solution was considered in each case.}
  \label{fig:validation_k}
\end{figure}

We ran \cloe{} for \( K \in \left\{ 3, 4, 5, 6 \right\} \) and performed model selection based on the log-posterior values of the MAP estimates. In every case we were able to identify the correct number of clones (Figure~\ref{fig:validation_k}), suggesting that either a moderate depth of sequencing or multiple samples should suffice in obtaining good estimates of the number of clones.

\paragraph{Reconstruction fidelity}

\begin{figure}
  \centering
  \makebox[\textwidth][c]{
    \includegraphics[scale=0.6]{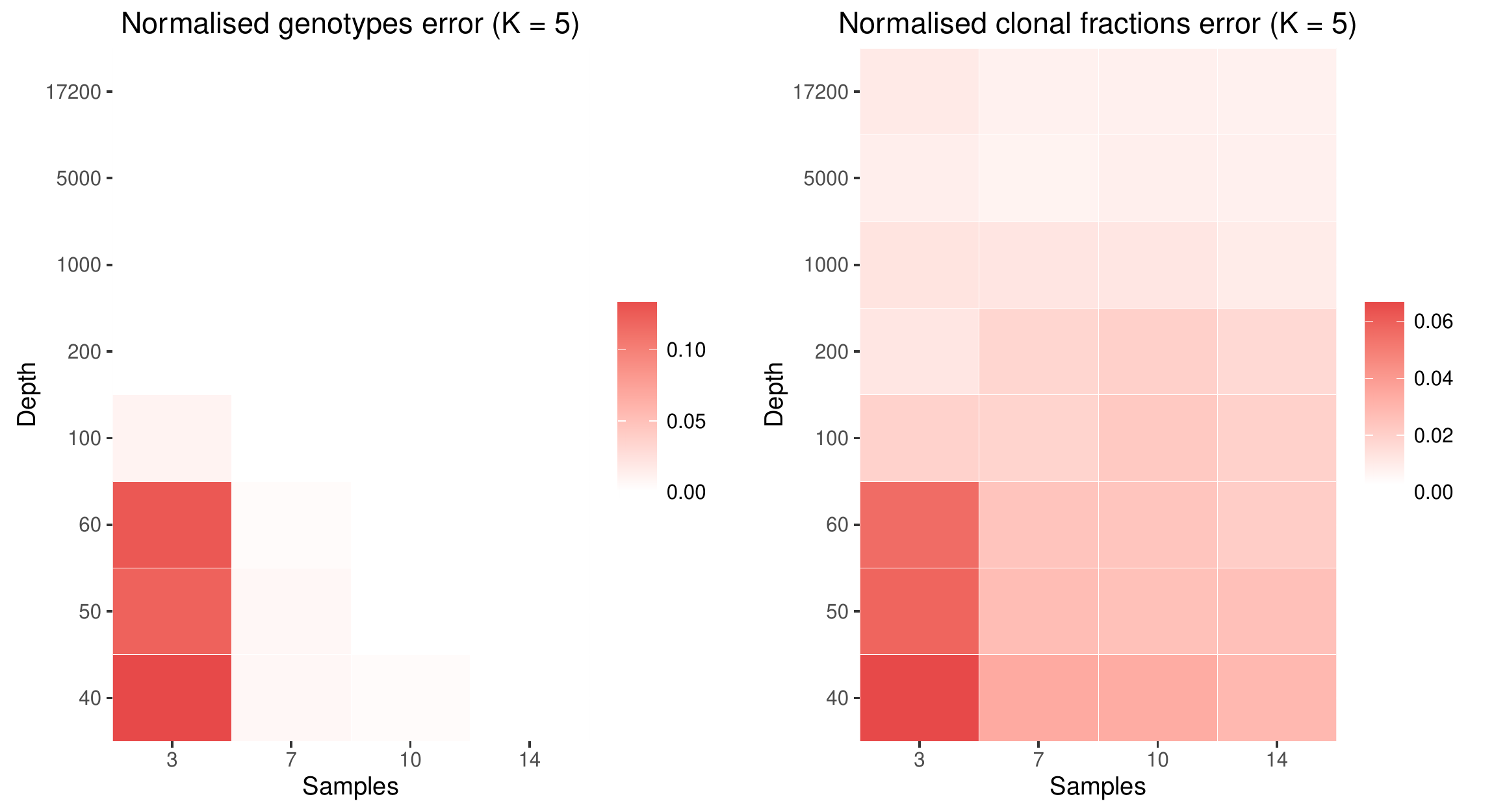}
  }
  \caption{Reconstruction errors on validation data obtained running \cloe{} with \( K = 5 \) clones. The heatmaps show the genotypes error (left) and clonal fractions error (right) for various combinations of depth and samples.}
  \label{fig:validation_err5}
\end{figure}

Overall, we obtained precise reconstructions for almost all depth-samples combinations. Considering for each combination only the first solution suggested by \cloe, on average 1\% of mutation assignments were inaccurate (\( Z_{err} \) median 0, mean 0.013), and clonal fractions were inferred with an average error lower than 2\% (\( F_{err} \) median 0.017, mean 0.019). As expected, we observed a pattern of decreasing errors as the data increase in the number of samples or in depth (Figure~\ref{fig:validation_err5}).

\paragraph{Specific low-depth cases}
Poorer reconstructions were obtained at lower depths (\( \leq 60\times \)) for the datasets with three samples. In every case, the inferred genotypes showed a faulty separation between two expected genotypes (Supplementary Figure 6), which led to high error metrics: \( Z_{err} \leq 0.134 \) and \( F_{err} \leq 0.065 \). Despite the imprecise reconstruction, there is an overall good agreement with the observed data (Supplementary Figure 6 (e)).

These results could be improved by tuning the running parameters of \cloe{} for these datasets. Because the height of the posterior peaks at these levels of depth is lower than at high depth, using less tempered chains may result in higher acceptance of chain swaps, and, consequently, in a more complete exploration of the posterior space. Increasing the number of MCMCMC iterations could also prove beneficial.

It should be also noted that at low depths sampling noise may promote suboptimal parameter combinations to near-optimal. In this case, more mutations should be analysed in order to average sampling noise effects, though this may place a heavy burden on our implementation's runtime. Alternatively, one could model more data in terms of samples. If the clonal fractions are dynamic enough, meaning that most clones grow and shrink at some point in the samples, more opportunities are provided to separate clonal signals.


\subsection{Comparison to previous approaches}
\label{sec:benchmark}

To further benchmark \cloe, we compared the results of three published methods compatible with targeted sequencing on our validation dataset: BayClone \cite{Sengupta2015} and Clomial \cite{Zare2014}, two latent feature models, and PyClone \cite{Roth2014}, a non-parametric model. Other methods, like PhyloWGS~\cite{Deshwar2015} or CloneHD~\cite{Fischer2014}, are not applicable to targeted sequencing.

\begin{figure}
  \centering
  \includegraphics[scale=0.56]{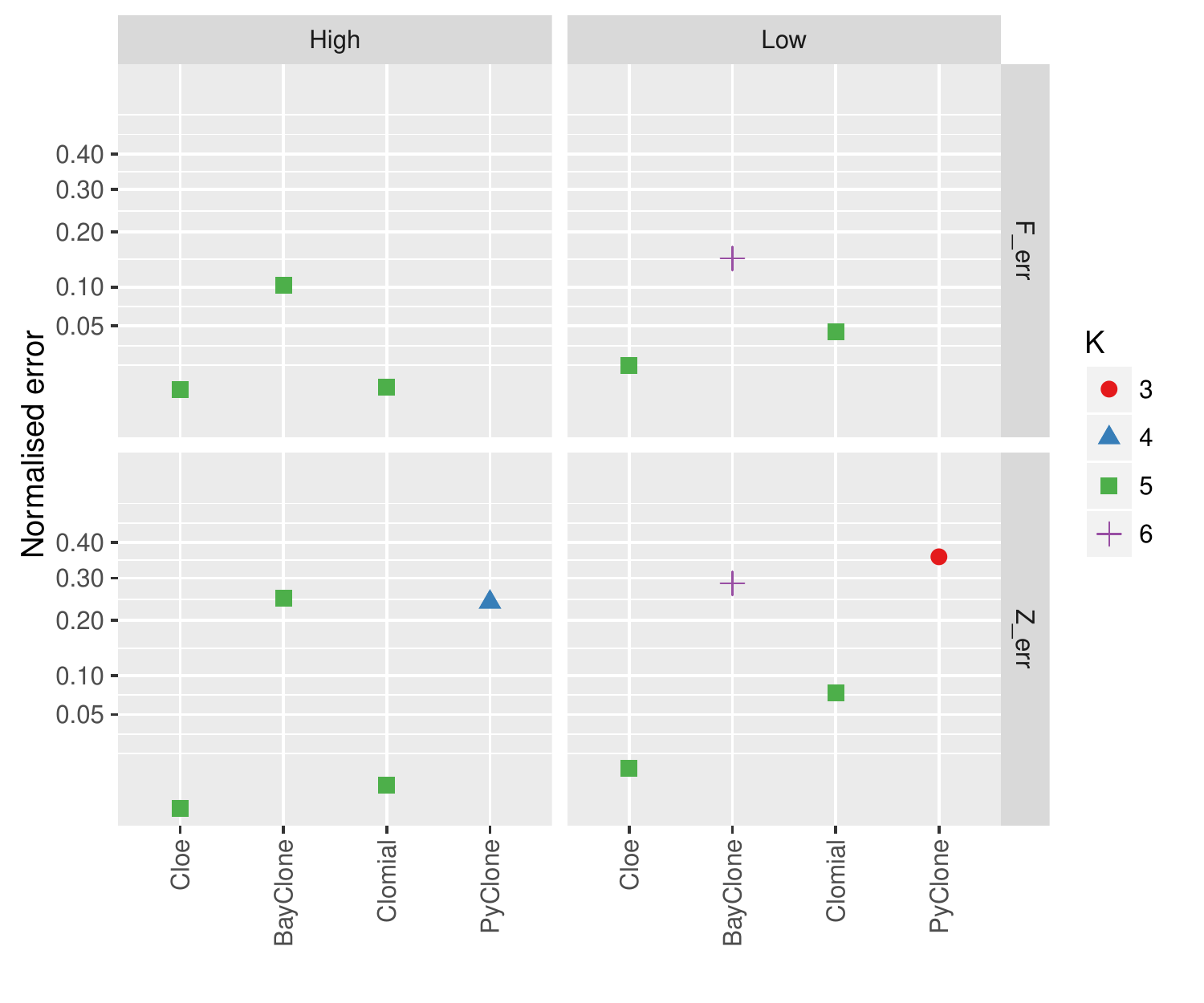}
  \caption{Comparison of \cloe{} against three published methods on our validation dataset, consisting of targeted sequencing for 82 mutations in 14 samples (average depth 17260\( \times \)); five clones are present in the data. \( Z_{err} \) denotes the reconstruction error on genotypes; \( F_{err} \) the error on clonal fractions. The two datasets (17260\( \times \) depth and 14 samples, and 100\( \times \) and 3 samples) are denoted ``high'' and ``low'', respectively. The legend refers to the number of inferred clones.}
  \label{fig:method_comparison}
\end{figure}

We ran two tests on the validation dataset described in the last section, first using all samples and the entire depth, and then 3 samples and a depth of 100\( \times \). Our method's performance on the first dataset is a perfect reconstruction of the genotypes (\(Z_{err} = 0 \)) and a near-perfect reconstruction of the clonal fractions (\(F_{err} = 0.005 \)), with a correct identification of the number of clones. With less data, there are three misassignment (\( Z_{err} = 0.009 \)) and the error of the clonal fractions is 0.017 (Figure~\ref{fig:method_comparison}); again, the number of clones is correctly inferred.

\begin{figure}
  \centering
  \makebox[\textwidth][c]{
    \includegraphics[scale=0.56]{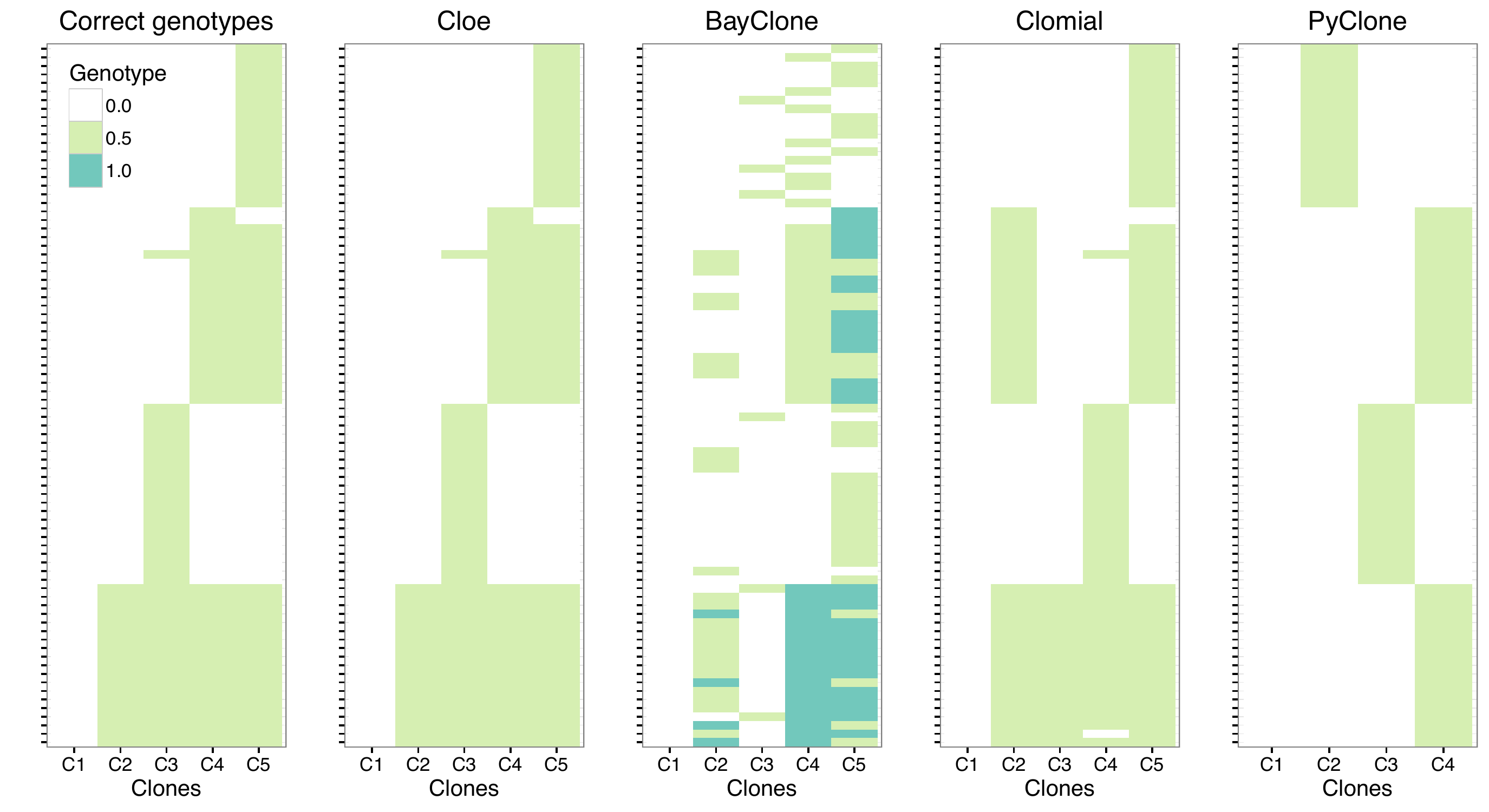}
  }
  \caption{Comparison of the genotypes inferred by the four benchmarked methods using all the data in our validation dataset. PyClone's reconstruction is a clustering of the mutations. In this representation of the genotypes we padded solutions with the normal clone (C1) for a more direct comparison with our method. The legend refers to the proportion of mutated alleles out of two.
  }
  \label{fig:method_comparison_z}
\end{figure}

BayClone \cite{bayclone} was run with default parameters for 45000 iterations, discarding the first 5000 and thinning the chain by a factor of 4. We tested the same model sizes as for our own method, namely 3, 4, 5, and 6. Through the log-pseudo-marginal likelihood, BayClone was able to identify \( K = 5 \) as the best solution on the first dataset. However, the reconstruction of the genotypes was less precise (Figure~\ref{fig:method_comparison_z}), with \( Z_{err} = 0.25 \) and \( F_{err} = 0.103 \). Here \( Z_{err} \), the normalised absolute difference between inferred and real genotypes matrices, ignores the ploidy of the mutation. The reconstruction was poorer on the second dataset because of the less precise data: six clones were inferred with \( Z_{err} = 0.287 \), and \( F_{err} = 0.147 \) (Supplementary Figure 7).

Clomial (version 1.6.0) implements an EM algorithm, and it was run with default parameters (1000 restarts, and 100 maximum EM iterations) using model sizes of 3, 4, 5, and 6. On the first dataset, model selection with BIC (and AIC) indicated \( K = 5 \) as the best solution, with one misassignment (\( Z_{err} = 0.003 \)) and an accurate reconstruction of the clonal fractions (\( F_{err} = 0.006 \)). With less data, model selection was unclear as AIC, BIC and the log-likelihoods were all discordant. Using the correct model size Clomial obtained a \( Z_{err} = 0.076 \) and \( F_{err} = 0.044 \).

PyClone (version 0.12.9) was run for 30000 iterations with a beta-binomial density and copy-number neutral states allowing a single mutant allele out of two (AB mode). PyClone was also provided with estimates of cellularity for each of the samples. We removed the first 3000 iterations as burn-in samples and thinned the chain by a factor of 4. The output of PyClone consists of a clustering of the observed mutations, 
where each cluster should roughly correspond to one of the non-root nodes of Figure~\ref{fig:data_cnn}. Phylogenetic modelling can translate these clusters into genotypes.
On the full dataset, PyClone produced three clusters, as shown in Figure~\ref{fig:method_comparison_z}. Because the cluster of stem mutations was merged with one of its two children, we were unable to interpret the results phylogenetically. Hence, we could not derive genotypes nor clonal fractions. The estimate of \( K \) is 4 with \( Z_{err} = 0.241 \). On less data, two clusters were produced, leading to an estimate of \( K \) of 3, with \( Z_{err} = 0.357 \).

In summary, our benchmark shows that \cloe{} compares favourably against similar published methods (Figure~\ref{fig:method_comparison_z}). It is expected that the accuracy of the reconstruction would be affected by the quality of the data. Indeed every model performed more poorly on less data, however \cloe{} seemed to be affected to a lesser extent (Supplementary Figure 7).

\section{Case studies}

We show the applicability of \cloe{} to clinical data in two case studies.


\subsection{Chronic lymphocytic leukaemia}
This dataset consists of five time points for each of three chronic lymphocytic leukaemia patients \cite{Schuh2012}. The original study identified mutations by whole-genome sequencing (WGS; average depth across the mutation loci 39\( \times \)) and quantified a subset of these with deep targeted sequencing (TAR; average depth 101600\( \times \)). 

The authors' analysis reported evolutionary trees and clonal fractions for each of the three cases. We used this information to run \cloe{} with known clonal fractions on all mutations, prioritising information from the higher depth datasets. We interpreted these results as ground truth mutation assignments for all three patients, and scored our reconstructions to these reference parameters.

We ran \cloe{} on the reported mutations with \( K \in \left\{ 3 \dots 7 \right\} \) for each case and each experiment (WGS, low-depth, and TAR, high-depth), comparing our results with the original study, with PhyloSub's results \cite{Jiao2014}, and with CloneHD's reconstruction of case CLL003 \cite{Fischer2014}. We also included another dataset, which consisted of the WGS dataset with data from the higher depth TAR dataset for mutations in common.

\begin{figure}
  \centering
  \includegraphics[scale=0.75]{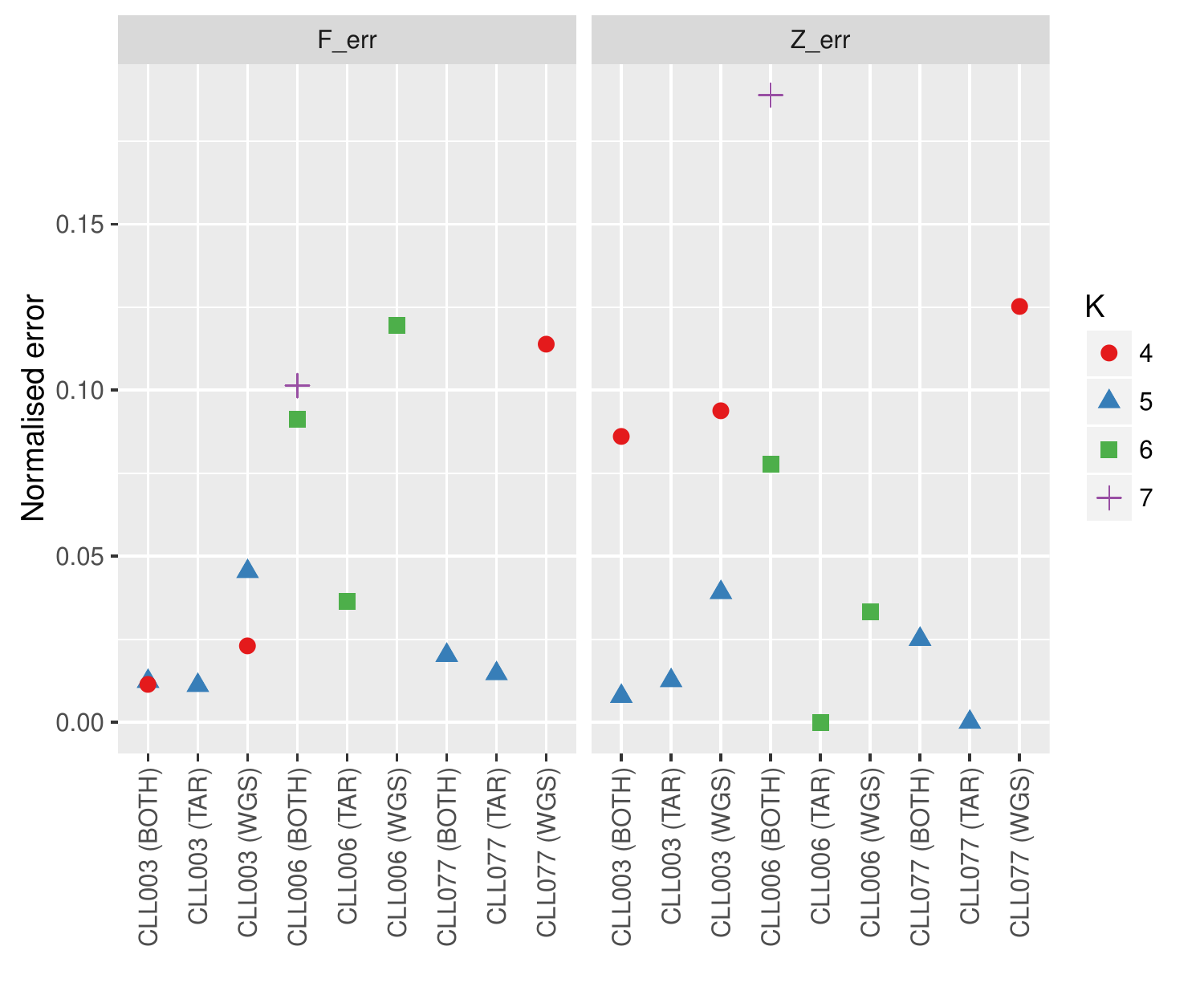}
  \caption{Performance metrics of \cloe{} on the CLL datasets. The correct number of clones for cases CLL003 and CLL077 is 5, whereas for CLL006 it is 6. TAR stands for targeted sequencing (average depth 101600\( \times \)); WGS stands for whole-genome sequencing (average depth 39\( \times \); BOTH is the WGS dataset with TAR data for shared mutations. The legend refers to the number of inferred clones. When the first solution inferred the wrong number of clones, the top solution for the correct number of clones is also shown.}
  \label{fig:cll_error}
\end{figure}

Case CLL003 displays a radical clonal shift (Supplementary Figure 8 (a) and (b)): the main clone in the early time points is replaced by a distinct new clone that appears only at the second time point and expands to become the predominant clone.
Using targeted sequencing data, \cloe{} obtained a very accurate reconstruction, identifying the correct number of clones, obtaining a single misassignment and average errors on clonal fractions of 1\% (Figure~\ref{fig:cll_error}, Supplementary Figure 9). On less data, our method opted for a solution with 4 clones that ignored the founding clone, only present in the first of five samples at a clonal fraction of 3\%. Choosing the top solution with 5 clones recovered the correct clonal structure; on WGS data there were five incorrect mutation assignments (Supplementary Figure 10), whereas with the combined dataset only one (Supplementary Figure 11). Barring the rare founding clone, the 4-clone reconstructions are correct with one (combined data) and two (WGS data) misassignments.

The remaining cases showed more stable dynamics (Supplementary Figure 8 (c)--(f)). For CLL006, \cloe{} assigned the nine mutations of the TAR dataset to six clones without errors; three errors were observed with 18 mutations in the WGS dataset (Figure~\ref{fig:cll_error}). Analysis of the combination of the two data types yielded an additional clone, though similar log-posterior probabilities and a higher log-likelihood were obtained by a six-clone solution. Removing clone C5 from the seven-clone solution yields a correct reconstruction (Supplementary Figure 12).

Finally, for CLL077, \cloe's analysis resulted in a perfect reconstruction of the genotypes with targeted sequencing data. Two misassignments were obtained for the combined dataset, whereas four of the five clones were identified in the WGS data: the founding clone, with only four of the 20 mutations, was merged with one of its children. After the four-clone solutions, solutions with six-clones had high log-posterior probabilities. Indeed the first of these solutions is an accurate reconstruction with two misassignments and one clone repeated twice almost identically. In the middle, solutions with the expected number of clones, five, had six errors (Supplementary Figure 13).

Overall, \cloe{} produced accurate reconstructions of the latent parameters. Higher errors were observed when an incorrect number of clones was inferred. However, even in these cases, our phylogenetic model allowed us to obtain close approximations of the ground truth. 

Assuming that our reconstruction of the ground truth is correct, \cloe's inference results in a better reconstruction than reported by CloneHD \cite{Fischer2014}, both using high-depth and low-depth data: first, because of our phylogenetic modelling, we were able to identify the founding clone; second, we could confidently identify the rising clone's parent (the ambiguous green clone in \cite{Fischer2014}). With low-depth data, \cloe{} did prefer a model with four clones, but could also provide a more accurate five-clone solution.

Our results on targeted sequencing data largely agree with those obtained by PhyloSub~\cite{Jiao2014}, with two small exceptions. For CLL003, \cloe{} predicts that clone 4 (clone \( c \) in \cite{Jiao2014}, Figure 7, right) does not harbour the \textit{IL11RA} mutation. This episode appears to be supported by the data (Supplementary Figure 14) as \cloe's reconstruction leads to closer fit with the data (sum of absolute errors on the allele fractions is 0.06 for \cloe, 0.13 for PhyloSub, for this mutation). Rather than a loss of mutation, this could be due to convergent evolution at the leaf nodes, leading to a sum of absolute errors of 0.07. For case CLL006, our reconstruction agrees with that of \cite{Schuh2012}: five tumour clones are detected, and the \textit{EGFR} mutation is predicted to stem from the founding clone. PhyloSub preferred to place the EGFR mutation in an additional clone after the founder, leading to a closer fit: the sum of absolute errors was 0.02 compared to \cloe's 0.07 for this mutation.


\subsection{Acute myeloid leukaemia}

AML31 refers to a patient with acute myeloid leukaemia, whose case was studied in great depth with several sequencing experiments targeting bulk DNA at various scales, RNA and also single cells \cite{Griffith2015}. As each layer of data refined the authors' understanding of the evolution of this tumour, seven clusters and driver mutations were identified. Integration of all sequencing data revealed over 1300 mutations curated in a ``platinum list''. The tumour genomes appeared to be devoid of copy-number aberrations. 

We considered a subset of platinum-list mutations for three datasets: ALLDNA (median depth of 1841\( \times \) for the primary tumour sample, 388\( \times \) for the relapse), TORRENT (median depths 41\( \times \) and 46.5\( \times \)), and WGS (median depths 323\( \times \) and 41\( \times \)). For each dataset, we selected a random subset of 250 mutations, halving the number of mutant reads for hemizygous mutations, and adding reported driver mutations.

\begin{figure}
  \centering
  \includegraphics[width=\textwidth]{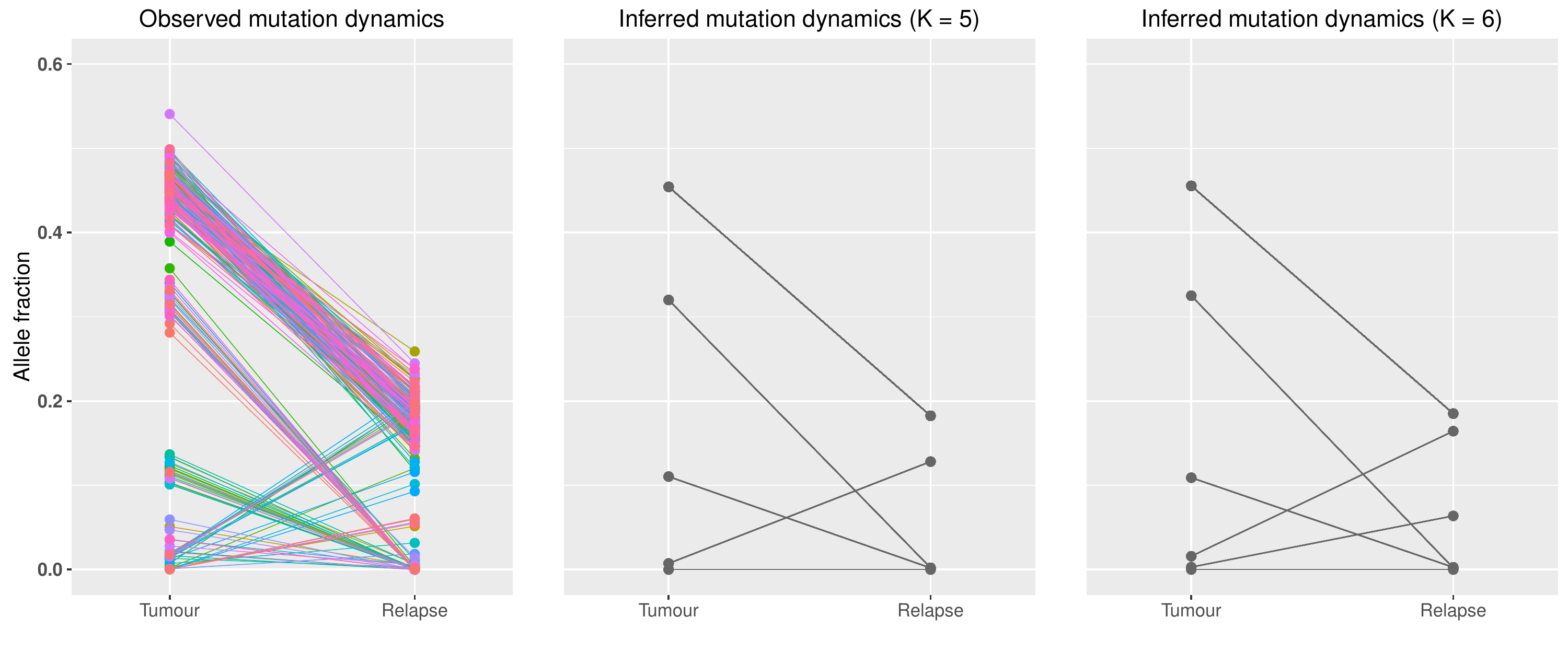}
  \caption{Observed and inferred mutation dynamics for 257 mutations from the ALLDNA dataset. Left: observed allele fractions; centre: allele fractions inferred by \cloe{} with 5 clones; right: allele fractions inferred by \cloe{} with 6 clones.}
  \label{fig:alldna_fit}
\end{figure}

\cloe{} was run with \( K \in \left\{ 3, \dots, 7 \right\} \) on the datasets. Model selection on ALLDNA indicated \( K = 5 \) as the preferred solution, followed closely by \( K = 6 \) which provided a closer fit to the data (Supplementary figure 15). The inferred mutation dynamics for both models are shown in Figure~\ref{fig:alldna_fit}. Whereas both model sizes could capture the trends in the data, the solution for \( K = 6 \) correctly identified two groups of mutations that rise in allele fraction in the relapse sample. 

\begin{figure}
  \centering
  \includegraphics[width=\textwidth]{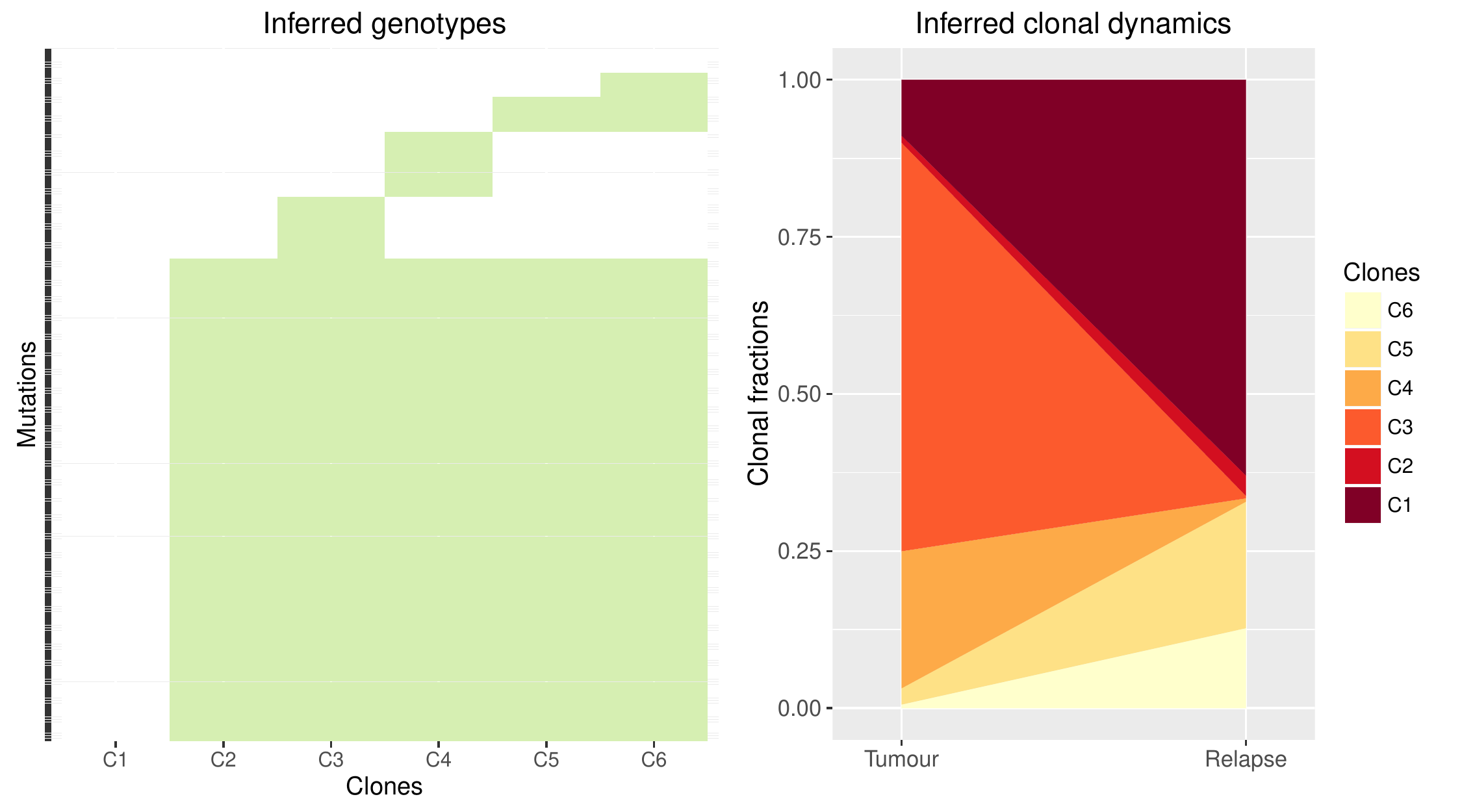}
  \caption{Parameters inferred by \cloe{} running with 6 clones on 257 mutations from the ALLDNA dataset. Genotypes are shown on the left, where green denotes presence of a mutation; clonal fractions for each clone are shown on the right. C1 is fixed as the normal contamination.}
  \label{fig:alldna_params}
\end{figure}

Our reconstruction shows a decrease in tumour burden at relapse, a single origin for all clones, and branched evolution after the founding clone (Figure~\ref{fig:alldna_params}). Clone 5 and its child, clone 6, become the main clones in the relapse sample, supplanting clones 3 and 4. The founding clone appears present only at very low clonal fractions.

Matching our clones to the original clusters, we found a close correspondence (Table~\ref{tab:alldna_match}), corroborating \cloe's inference. The only misassignment is \textit{TP53} to clone 6, which in the original study required single-cell sequencing and additional time points to identify as belonging to a separate clone. Beyond the genotypes, there was also a close match between inferred and expected clonal fractions, with a maximum absolute difference of 4\%.

Cluster 6 was not identified by our model. According to the original analysis, this cluster was present at less than 5.5\% clonal fraction in the primary sample, to then disappear at relapse. Such a cluster would contribute half of its clonal fraction in allele fraction, due to the heterozygosity of the mutations. We do observe that nine of the 257 mutations were not assigned to any clone. Their average allele fraction was 2.2\% in the primary sample and close to 0\% in the relapse (Supplementary Figure 16). Because they did not fit the dynamics of the other clones, sequencing noise was used to fit them (eq.~\ref{eq:seq_noise}). 

\begin{table}
  \centering
  \begin{tabular}{ccl}
\hline
Clone & Cluster & Drivers \\
\hline
C2 & 1 & \textit{DNMT3A} \\
C4 & 4 & \textit{FOXP1} \\
C5 & 3 & \textit{IDH2} \\
C3 & 2 & \textit{IDH1} \\
C6 & 5 & \textit{CXCL17}, \textit{TP53} \\
\hline
  \end{tabular}
  \caption{Correspondence between \cloe's inferred clones, and the clusters in the original analysis by \cite{Griffith2015}. While drivers are also present in the children of a clone, here we report the clone in which the mutations first appeared.}
  \label{tab:alldna_match}
\end{table}

\begin{figure}
  \centering
  \includegraphics[width=\textwidth]{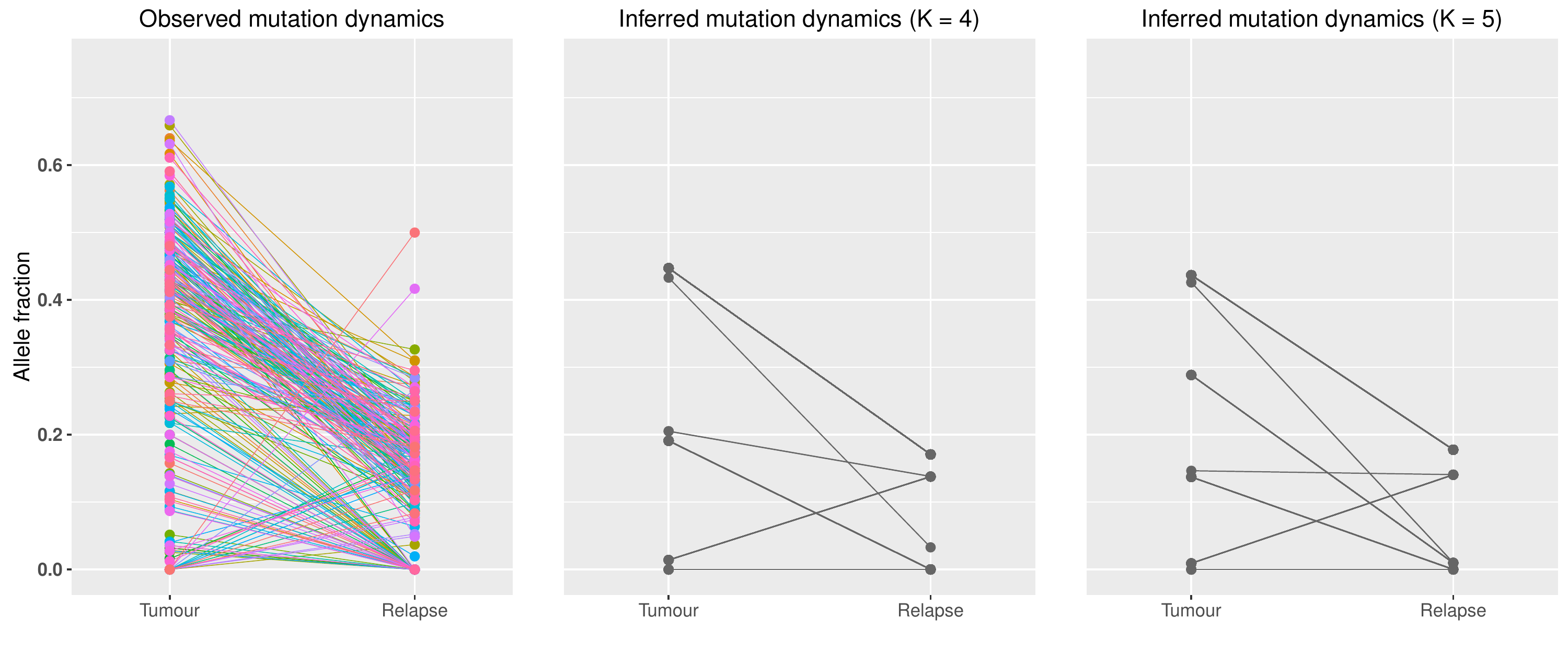}
  \caption{Observed and inferred mutation dynamics for 254 mutations from the TORRENT dataset. Left: observed allele fractions; centre: allele fractions inferred by \cloe{} with 4 clones; right: allele fractions inferred by \cloe{} with 5 clones.}
  \label{fig:torrent_fit}
\end{figure}

On the modest amount of data of the TORRENT dataset our model selection produced a more conservative estimate of the number of clones, preferring four clones. Using five or more clones improved the log-likelihood to the same extent. We compare here solutions for \( K = 4 \) and \( K = 5 \) (Figure~\ref{fig:torrent_fit}).

With three tumour clones, our model matched the main trends: two large clones in the primary sample that disappear at relapse, and one growing clone. In addition, tumour content was accurately inferred: 89\% for the primary sample and 44\% for the relapse sample, compared to the expected values of 91\% and 47\%. The addition of a fourth tumour clone (\( K = 5 \)) allows a better disambiguation of the clones present in the primary, while the spread of allele fractions in the relapse sample makes it difficult to distinguish two rising clone.

Identifying seven clones including the normal in two samples with a median depth less than 45\( \times \) is an arduous task. \cite{Griffith2015} show that SciClone detects four clones up to around 100\( \times \) depth using all mutations on the platinum list. While \cloe{} prefers four clones using a subset of mutations at a depth of 45\( \times \), it is capable of splitting the observed dynamics further, obtaining closer approximations of the real clonal structure.

Finally, for the WGS dataset, \cloe's solution with 5 clones obtained the highest posterior probability, while 6 and 7 clones obtained closer fits to the data (Supplementary Figure 17). With four tumour clones, \cloe{} identified three decreasing groups of mutations and one group that arose at relapse. This matches the observed dynamics, as the low depth at relapse accounts for a larger spread of the allele fractions that confounds the identification of two rising clones (Supplementary figure 18). Interestingly, the addition of another clone, rather than fitting this low-depth relapse data, matches a fourth group of mutations present only in the primary around 5\% clonal fraction. These mutations overlap with the unassigned mutations in the ALLDNA dataset and the inferred clone does not harbour additional driver mutations other than \textit{DNMT3A}, which derives from its parent.

With this case study we applied \cloe{} to a scenario with two samples, highlighting the difficulties of automatic model selection, especially when trying to identify a large number of clones with a moderate amount of data.

The running parameters for the two case studies differ from the ones listed in Table~\ref{tab:params} in that we used five chains with \( \Delta T = 0.25 \). In addition, for the AML datasets we ran 50000 iterations of our sampler with \( \mu = 0.2 \), \( \rho = 0.04 \), and \( \varepsilon = 0.001 \).


\section{Discussion}

As tumour sequencing data grow in depth and breadth, the question of tumour heterogeneity will continue to be focal. In this study we presented \cloe, a novel latent feature model for direct clonal reconstruction. Our model discovers genotypes in the data by assigning observed mutations to latent features (clones) guided by a latent phylogeny. This phylogenetic deconvolution sets \cloe{} apart from other direct reconstruction methods \cite{Fischer2014, Zare2014, Sengupta2015}. Compared to indirect reconstruction methods, our algorithm can handle multiple primary tumours, the loss of mutations and convergent evolution. In particular, to our knowledge this is the first method to allow and penalise convergent evolution.

Our study on simulated data showed a good performance of our MCMCMC algorithm. However, tuning the MCMCMC parameters in order to correctly explore the spiked posterior landscape is not trivial. We empirically found parameters that would allow the chains to mix well. Regions of high posterior probability are quickly reached, yet finding the right peak is a slow process, complicated by each biological constraint on the model.
Many parameters can be tuned in our model. We sought values that would work well for both simulated data and our validation data. Tuning the MCMCMC parameters to each dataset independently, thus optimising the exploration of the posterior space, might further improve results.

In our definition of the tree we assume that multiple primary tumours are less likely to occur than tumours with a single origin. If our understanding of clonal evolution were to suggest otherwise, the definition of the tree may be simplified to a discrete uniform distribution, giving equal weight to a single origin or multiple ones.

\paragraph{Limitations}
The main limitation of our method is the restriction to mutations from copy-number neutral regions. Whereas this may be amenable to certain types of cancer (e.g. mutation-driven rather than copy-number driven cancers), it may preclude the analysis of more genomically rearranged tumours.

In contrast to some models described in the literature, our method does not include the number of clones as a parameter. Instead, \cloe{} must be run for various choices of \( K \), and the best solution in terms of posterior probability will indicate the number of clones with good accuracy. On our simulation and validation datasets our model was indeed able to identify the correct number of clones in 58/59 cases. 

As shown in the case studies, model selection may not be trivial. We thus recommend manual review of the inferred parameters for various model sizes to ensure that the results of the inference are robust.

Analysing hundreds of mutations can result in a high computational burden. This limitation could be alleviated by preprocessing the input data, grouping mutations that exhibit similar dynamics throughout the samples. One way to do this is via a Chinese Restaurant Process with a product of binomials; mutant read counts and depths for all mutations in a cluster could then be summed and analysed as a single unit.

\paragraph{Extensions}
We see several avenues for future extensions.
At the theoretical level, future work should focus on optimising the inference and extending this framework to arbitrary copy-numbers. Also, to address the model selection problem, the phylogenetic latent feature model could be rephrased in a non-parametric perspective.
In terms of applications, our model could also be applied to epigenetics: by appropriately changing the likelihood function, \cloe{} could deconvolute methylation data into evolutionarily related epigenotypes.

In summary, \cloe{} is a rigorous and flexible framework for clonal deconvolution of cancer genomes that achieves high accuracy in benchmarking studies and leads to important insights into tumour evolution in clinical case studies.


\section*{Acknowledgements}

We would like to acknowledge the support of The University of Cambridge, Cancer Research UK and Hutchison Whampoa Limited. 
This work was funded by CRUK core grant C14303/A17197, in particular A19274 (Markowetz lab core grant).
We wish to thank Marta Grzelak and James Hadfield for their assistance with sequencing, and Malvina Josephidou for discussions on the CRP clustering of mutations.

\bibliographystyle{plain}
\bibliography{heterogeneity,ctdna,methods}

\begin{supplement}
  \sname{Supplement A}
  \label{suppA}
  \stitle{Supplementary figures}
  \slink[url]{./supplementary\_figures.pdf}
  \sdatatype{.pdf}
  \sdescription{Supplementary figures.}
\end{supplement}

\begin{supplement}
  \sname{Supplement B}
  \label{suppB}
  \stitle{Source code of the analyses}
  \slink[url]{./reproduce.tar.gz}
  \sdatatype{.tar.gz}
  \sdescription{This package contains scripts, data (in the form of matrices of mutant read counts and depths) analysed in this article, and a version of \cloe{} to reproduce the findings.}
\end{supplement}

\end{document}